\documentclass[preprintnumbers, prd, twocolumn, showpacs, floatfix, preprintnumbers, 
letterpaper, 
superscriptaddress,nofootinbib]{revtex4-1}
\pdfoutput=1 
\usepackage{graphicx,epsfig}
\usepackage {amssymb}
\usepackage {longtable}
\usepackage{multirow}
 \usepackage{appendix}
 \usepackage{dcolumn}
\usepackage{bm}
\usepackage{amsfonts}
\usepackage{subfigure}
\usepackage{color}
\usepackage{relsize}
\usepackage{relsize}
\usepackage{amsmath}
\usepackage{xparse}
\usepackage{accents}

\ExplSyntaxOn\makeatletter
\renewcommand*\dddot[1]{%
  \placeaccent{\acc@dot\mkern1.4mu\acc@dot\mkern1.4mu\acc@dot}{#1}%
  }
\renewcommand*\ddddot[1]{%

\placeaccent{\acc@dot\mkern1.4mu\acc@dot\mkern1.4mu\acc@dot\mkern1.4mu\acc@dot}{#1}%
  }
\NewDocumentCommand \vardot {O{1} m }
  {
    \int_compare:nNnTF
      {#1} = {1}
      {\dot #2}
      {\placeaccent{\prg_replicate:nn {#1-1} {\acc@dot\mkern1.4mu}\acc@dot}{#2}}
  }
\newcommand*\placeaccent[2]{%
  \begingroup
  \def\acc@dot{\kern-0.08em.\kern-0.08em}%
  \def\acc@skip{\ifx\macc@style\displaystyle0.32
           \else\ifx\macc@style\textstyle0.32
           \else\ifx\macc@style\scriptstyle0.22
           \else0.15\fi\fi\fi ex}%
  \def\mathaccent##1##2{%
    \setbox6\hbox{$\m@th\macc@style#1$}%
    \@tempdima\wd4
    \advance\@tempdima\macc@kerna
    \advance\@tempdima-\wd6
    \divide\@tempdima\tw@
    \@tempdimb\z@
    \ifdim\@tempdima<\z@ \@tempdimb-\@tempdima \@tempdima\z@ \fi
    \vbox{\offinterlineskip
          \moveright\@tempdima\box6
          \kern\acc@skip
          \moveright\@tempdimb\box4}%
  }%
  \macc@depth\@ne
  \let\math@bgroup\@empty \let\math@egroup\macc@set@skewchar
  \mathsurround\z@ \frozen@everymath{\mathgroup\macc@group\relax}%
  \macc@set@skewchar\relax
  \let\mathaccentV\macc@nested@a
  \macc@nested@a\relax111{#2}%
  \endgroup
}
\makeatother\ExplSyntaxOff

\newcommand{\be}{\begin{equation}}
\newcommand{\ee}{\end{equation}}
\newcommand{\bea}{\begin{eqnarray}}
\newcommand{\eea}{\end{eqnarray}}

\begin{document}

\title{Modified teleparallel gravity with higher-derivative torsion terms}

\author{Giovanni Otalora}
\email{giovanni@ift.unesp.br}
\affiliation{Departamento de Matem\'atica, ICE, Universidade Federal de Juiz
de Fora, Minas Gerais, Brazil}\affiliation{Instituto de F\'isica Te\'orica, 
UNESP-Universidade 
Estadual Paulista
Caixa Postal 70532-2, 01156-970, S\~ao Paulo, Brazil}
\affiliation{Instituto de F\'{\i}sica, Pontificia Universidad Cat\'olica de 
Valpara\'{\i}so, 
Casilla 4950, Valpara\'{\i}so, Chile}
 
\author{Emmanuel N. Saridakis}
\email{Emmanuel\_Saridakis@baylor.edu}
\affiliation{Instituto de F\'{\i}sica, Pontificia Universidad Cat\'olica de 
Valpara\'{\i}so, 
Casilla 4950, Valpara\'{\i}so, Chile}
\affiliation{Physics Division,
National Technical University of Athens, 15780 Zografou Campus,
Athens, Greece} 
\affiliation{CASPER, Physics Department, Baylor University, Waco, TX 76798-7310, USA}

\begin{abstract} 
We  construct $F(T,\left(\nabla{T}\right)^2,\Box {T})$  gravitational modifications, 
which are  novel classes of modified theories arising from  higher-derivative torsional 
terms in the action, and are different than their curvature analogue. Applying  them in a 
cosmological framework we obtain an effective dark energy sector that comprises of the 
novel torsional contributions. We perform a detailed dynamical analysis for two specific 
examples, extracting the stable late-time solutions and calculating the corresponding 
observables. We show that the thermal history of the universe can be reproduced,
and it can result in a dark-energy dominated, accelerating universe, where the 
dark-energy 
equation-of-state parameter lies in the quintessence regime, or may exhibit 
the phantom-divide crossing during the cosmological evolution. Finally, the scale factor 
behaves asymptotically either as a power-law or as an exponential, 
in agreement with observations.
\end{abstract}


\maketitle

\section{Introduction}\label{Introduction}
The early and late time accelerated expansions of the universe are probably the most 
surprising findings in modern cosmology and establish a serious challenge to our current 
knowledge of physics. There are two main ways that one could follow in order to describe 
them. The first is to maintain general relativity as the gravitational theory and modify 
the content of the universe by introducing new, exotic components, such as the 
inflaton field(s) \cite{Inflation} or the dark energy sector
\cite{Copeland:2006wr,Amendola,Cai:2009zp}. The second is to modify the gravitational 
theory itself, constructing a theory with additional degrees of freedom that can drive 
acceleration, but which still possesses general relativity as a 
particular limit \cite{modgrav1}. 

Most works in modified gravity start from the standard gravitational formulation, which 
is based on curvature, and modify the Einstein-Hilbert action, with the 
simplest extended model being the $F(R)$ one \cite{F-R}. Also, other modifications to 
gravity can 
arise from a Planck-scale deformed dispersion relation and effective spacetime metric 
that 
depends 
of the energy, momentum or spin of the probe particle \cite{DeSitterR, 
DoublySpecialR,RainbowGravity, Deriglazov:2015bqa,Deriglazov:2015kba,Ramirez:2013xga}. 
Nevertheless, one can equally 
well build modified gravitational theories starting from the torsional 
gravitational formulation, and in particular from the Teleparallel Equivalent of General 
Relativity (TEGR) \cite{ein28,Hayashi79,Pereira.book,Maluf:2013gaa}. Since in this theory
the gravitational Lagrangian is the torsion scalar $T$, the simplest extended scenario is 
to extend it to $F(T)$ theory \cite{Bengochea:2008gz,Linder:2010py} (see 
\cite{Cai:2015emx} for a review). Note that although at the level of equations TEGR is 
completely equivalent with general relativity, $F(T)$ is a different class of modified 
gravity than $F(R)$ gravity, and therefore its cosmological implications bring novel 
features, either at late-times \cite{Dent:2011zz,Geng:2011aj,Bamba:2013jqa} or at the 
inflationary epoch\cite{F-T-Inf}.

 Nevertheless, in curvature-based modified gravity one can construct more complicated 
extensions of the Einstein-Hilbert action by introducing higher-order terms, such as 
 $R_{\mu\nu}R^{\mu\nu}$, $R_{\mu\nu\alpha\beta}R^{\mu\nu\alpha\beta}$, $\left(\nabla 
R\right)^2$, the Gauss-Bonnet 
combination, $R\Box R$, $R\Box^{k} R$, etc, and moreover couple these terms to 
an additional scalar field and its derivatives \cite{modgrav1}, since such terms could be 
justified due to quantum corrections or through a fundamental gravitational theory (for 
instance  such terms appear in the string effective Lagrangian  or in 
Kaluza-Klein theories, when the mechanism of 
dimensional reduction is used \cite{modgrav1,EffecLagrangian}), or in  
quantum-gravity-like effective actions   at scales closed to the
Planck one \cite{Vilkovisky}.  
In principle, one could follow the same direction in torsional gravity, i.e construct 
gravitational modifications using higher-order torsional terms. For instance, one could 
construct the teleparallel equivalent of the Gauss-Bonnet combination and insert 
arbitrary 
functions of it in the Lagrangian \cite{Kofinas:2014owa}, or extend the procedure to the 
teleparallel equivalent of Lovelock gravity \cite{Gonzalez:2015sha}.

In this work, and inspired by the corresponding curvature based modification  
\cite{Naruko:2015zze}, we are interested in constructing novel torsional gravitational 
modifications using higher-derivative, $\left(\nabla T\right)^2$ and $\Box{T}$ terms, i.e 
theories that are characterized by the Lagrangian $F(T,\left(\nabla T\right)^2,\Box{T})$, 
and 
investigate 
their cosmological implications. The plan of the work is as follows: In Section 
\ref{model} we construct $F(T,\left(\nabla T\right)^2,\Box{T})$ gravity and we apply it 
in a cosmological framework, extracting the cosmological equations and calculating 
various observables. In Section \ref{CosmDynamics} we analyze in detail two 
specific models, performing a dynamical analysis in order to reveal the global features 
of the corresponding cosmological behavior. Finally, in Section \ref{Conclusions} we 
summarize the obtained results. 

\section{The model}
\label{model}
In this section we construct modified teleparallel gravitational theories with higher 
derivative torsion contributions, extracting the general field equations, and we apply 
them in a cosmological framework.

\subsection{$F(T,\left(\nabla{T}\right)^2,\Box {T})$ gravity}

In teleparallel gravity the dynamical field is the vierbein $e^\mu_A$, which forms an 
orthonormal base for the tangent space at each point of a manifold. It is related to the 
metric through
\begin{equation}
g_{\mu\nu}=\eta_{A B} e^A_\mu 
e^B_\nu,
\label{metricvierbeinrel}
\end{equation}
 where greek indice span the coordinate space and latin indices span the 
tangent space. Additionally, one introduces the 
Weitzenb{\"{o}}ck connection  \cite{Pereira.book} 
$\overset{\mathbf{w}}{\Gamma}^\lambda_{\nu\mu}\equiv e^\lambda_A\:\partial_\mu e^A_\nu$,
and therefore the gravitational field is 
described by the torsion tensor
\begin{equation}
T^\rho_{\verb| |\mu\nu} \equiv e^\rho_A
\left( \partial_\mu e^A_\nu - \partial_\nu e^A_\mu \right).
\end{equation}
Hence, the Lagrangian of the theory is the torsion scalar $T$, constructed by
contractions of the torsion tensor as \cite{Maluf:2013gaa}
\begin{equation}
\label{Tdefscalar}
T\equiv\frac{1}{4}
T^{\rho \mu \nu}
T_{\rho \mu \nu}
+\frac{1}{2}T^{\rho \mu \nu }T_{\nu \mu\rho}
-T_{\rho \mu}{}^{\rho }T^{\nu\mu}{}_{\nu}.
\end{equation}

In the simplest torsional modified gravity, and inspired by similar procedures in 
curvature gravity,  one extends the Lagrangian to an arbitrary function $F(T)$, resulting 
to $F(T)$ gravity, \cite{Bengochea:2008gz,Linder:2010py}. However, one could be inspired 
by the higher-derivative curvature  modifications, and construct torsional modified 
gravity using higher derivative torsional terms, like  $\left(\nabla T\right)^2$ and 
$\Box{T}$. Hence, in this work we consider theories of the form 
\begin{equation}
S=\frac{1}{2}\int{d^{4}x e 
F(T,\left(\nabla{T}\right)^2,\Box {T})}+S_{m}(e^{A}_{\rho},\Psi_{m}), 
\label{action0}
\end{equation}
where $  \left(\nabla{T}\right)^2=\eta^{A B} e_A^\mu 
e_B^\nu \nabla_{\mu}{T}\nabla_{\nu}{T}=g^{\mu\nu} 
\nabla_{\mu}{T}\nabla_{\nu}{T}$ 
and $  \Box 
{T}= \eta^{A B} e_A^\mu 
e_B^\nu  \nabla_{\mu}\nabla_{\nu}{T}=g^{\mu\nu}\nabla_{\mu}{\nabla_{\nu}}{T}$, and 
where $e= \det \left(e^A_\mu \right)=\sqrt{-g}$  (for simplicity  we have set the light 
speed  $c=1$ and the gravitational constant $\kappa^{2}=8 \pi G=1$). Note that in the 
above total action we have also considered a general matter action comprised of general 
fields $\Psi_{m}$, allowing also for an arbitrary coupling with the vierbein. Finally, 
we mention that for simplicity in the present work we follow the usual, ``pure-tetrad'', 
approach to torsional modified gravity, while the extension to the inclusion of a general 
spin connection is straightforward, following \cite{Krssak:2015oua}.

Using for simplicity the notation   $X_{1}\equiv \left(\nabla{T}\right)^2$ and 
$X_{2}\equiv \Box {T}$,  as well as $F_{T}\equiv \partial F/\partial T$, $F_{X_{a}}\equiv 
\partial F/\partial X_{a}$, with $a=1,2$, variation of action (\ref{action0}) with 
respect to the vierbein leads to the following field equations: 
\begin{eqnarray}
&& 
\!\!\!
\frac{1}{e}\partial_{\mu}\left(e F_{T} e_{A}^{~\tau} S_{\tau}^{~\rho \mu}\right)-F_{T} 
e_{A}^{~\tau} S_{\nu}^{~\mu \rho} T^{\nu}_{~\mu \tau}+\frac{1}{4} 
e_{A}^{~\rho}F
\nonumber\\
&&
\!\!\!
+
\frac{1}{4}\sum_{a=1}^{2}\Bigg\{F_{X_{a}} \frac{\partial X_{a}}{\partial 
e^{A}_{~\rho}}\nonumber \\
&& \ \ \ \ 
-\frac{1}{
e}\Bigg[\partial_{\mu} \Bigg(e F_{X_{a}} 
\frac{\partial{X_{a}}}{\partial{\partial_{\mu}{e^{A}_{~\rho}}}}\Bigg)-
 \partial_{\mu}\partial_{\nu}\Bigg( e F_{X_{a}} \frac{\partial 
{X_{a}}}{\partial{\partial_{\mu}\partial_{\nu} e^{A}_{~\rho}}}\Bigg)\Bigg]\Bigg\}
\nonumber\\
&&\!\!\!
- 
\frac{1}{4 e}\partial_{\lambda}\partial_{\mu}\partial_{\nu} \Bigg(e F_{X_{2}} 
\frac{\partial{X_{2}}}{\partial_{\lambda}\partial_{\mu}\partial_{\nu}{e^{A}_{~\rho}}}
\Bigg)\nonumber \\
&&
\!\!\!
=\frac{1}{2} e_{A}^{~\tau}\,
  {\mathcal{T}^{(m)}}_{\tau}^{~\rho},
\label{FEquations}
\end{eqnarray}
where we have defined the ``superpotential'' $S_{\rho}^{~\mu\nu}\equiv 
\frac{1}{2}\left(K^{\mu\nu}_{~~\rho}+\delta^{\mu}_{\rho}\,T^{\theta\nu}_{
~~\theta}-\delta^{\nu}_{\rho}\,T^{\theta\mu}_{~~\theta}\right)$, with 
$K^{\mu\nu}_{~~\rho}\equiv 
-\frac{1}{2}\left(T^{\mu\nu}_{~~\rho}-T^{\nu\mu}_{~~\rho}-T_{\rho}^{~\mu\nu}\right)$   
the contortion tensor. Note that in the right hand side of  (\ref{FEquations})  we have 
defined the matter energy 
momentum tensor as
\begin{equation}
e_{A}^{~\tau}\,
  {\mathcal{T}^{(m)}}_{\tau}^{~\rho}
    \equiv   -\frac{1}{e} 
\frac{\delta{{\mathcal S}_m}}{\delta{e^{A}_{\rho}}}.
 \label{4}
\end{equation}
Finally, since the covariant derivative of the matter energy-momentum tensor in every 
theory where matter is minimally coupled to gravity is zero, as long as the matter 
Lagrangian is diffeomorphism invariant \cite{Weinberg:1972kfs}, we deduce that if in 
action  (\ref{action0}) we make the usual consideration that the matter action does not 
have an arbitrary coupling with gravity (i.e with the vierbein) but only a minimal 
coupling, then its covariant derivative is indeed zero. This can be verified explicitly 
too, by taking in this case the covariant derivative of  (\ref{FEquations}).
  
Equations (\ref{FEquations}) contain higher-order derivatives as expected. However, this 
is not necessarily an indication of Ostrogradsky instabilities 
\cite{Ostrogradksi50,Woodard:2015zca} since the above modified gravity has not been 
formulated in the Einstein frame. Thus, the higher-order derivatives may be just an 
indication of extra degrees of freedom, as it is the case in many gravitational 
modifications, like $f(R)$ gravity \cite{modgrav1}. One could try to transform the model 
in the Einstein frame, however in torsional modified gravities such transformations do 
not exist, or at least they are not known yet \cite{Yang:2010ji,Cai:2015emx}. Hence, the 
only safe method to examine whether the present constructions have any ghost or Laplacian 
instabilities, or extract the sub-classes that are free of such instabilities, is through 
a robust Hamiltonian analysis. Such a necessary investigation  lies beyond the scope of 
the present work, which is a first study on the subject, and hence it is left for a 
separate project.
 
\subsection{Cosmological equations}

In order to proceed to the cosmological applications of the above theory, we consider a 
flat Friedmann-Robertson-Walker (FRW) background space-time 
with metric $ds^2= dt^2-a^2(t)\,\delta_{ij} dx^i dx^j$, which arises from the vierbein
\begin{equation}
\label{weproudlyuse}
e_{\mu}^A={\rm
diag}(1,a(t),a(t),a(t)),
\end{equation}
where $a(t)$ is the scale factor. In such a geometry, and assuming as usual 
that the matter action includes only a minimal coupling to gravity (i.e with the 
vierbein),
the field equations 
(\ref{FEquations}) give rise to the two Friedmann equations as
\begin{eqnarray}
\label{Fr1}
&& F_{T} 
H^2+\left(24{{H}^{2}}{{F}_{X_{1}}}+{{F}_{X_{2}}}\right)\left(3 H 
\dot{H}+\ddot{H}\right)H\nonumber
\\
&&
+{{F}_{X_{2}}}\dot{H}^{2}
+
\left( 3 H^{2}-\dot{H}\right)H\dot{F}_{X_{2}}+24 
H^{3}\dot{H}\dot{F}_{X_{1}}\nonumber
\\
&&
+H^{2}\ddot{F}_{X_{2}}+\frac{F}{12}=\frac{\rho_{m}}{6},
\end{eqnarray}
\begin{eqnarray}
&& {F}_{T} \dot{H}+ H \dot{F}_{T}+24 H \left[2 H \ddot{H} +3\left( 
\dot{H}+H^{2}\right)\dot{H} \right]\dot{F}_{X_{1}}
\nonumber
\\
&&
+12 H \dot{H} \dot{F}_{X_{2}}+24 
H^{2}\dot{H}\ddot{F}_{X_{1}}+\left(\dot{H} +3 
H^{2}\right)\ddot{F}_{X_{2}}\nonumber
\\
&&
+24 H^{2} {F}_{X_{1}} \dddot{H}+H\dddot{F}_{X_{2}}+24 {{F}_{X_{1}}} {{\dot{H}}^{2}}\left( 
12 {{H}^{2}}+\dot{H}\right)\nonumber \\
&&
+24 
H {F}_{X_
{1}}\left( 4 \dot{H} +
3 {{H}^{2}}\right)\ddot{H}=-\frac{p_{m}}{2},
\label{Fr2}
\end{eqnarray}
where
\begin{eqnarray}
&&\!\!\!\!\!\!\!\!\!\! 
\dot{F}_{T}=F_{T T} \dot{T}+\sum_{a=1}^{2} F_{T X_{a}}\dot{X_{a}}.\\
&&\!\!\!\!\!\!\!\!\!\! 
\dot{F}_{X_{a}}=F_{X_{a} T} 
\dot{T}+\sum_{b=1}^{2}{F_{X_{a} X_{b}}\dot{X_{b}}}.\\
&&\!\!\!\!\!\!\!\!\!\! 
\ddot{F}_{X_{a}}=\left[F_{X_{a} T T} 
\dot{T}+\sum_{b=1}^{2}{F_{X_{a} T 
X_{b}}\dot{X_{b}}}\right]\dot{T}
\nonumber\\
&&\ \ \ \ \, +
\sum_{b=1}^{2}{\left[F_{X_{a}X_{b} T}\dot{T}+\sum_{c=1}^{2}{F_{X_{a} X_{b} 
X_{c}}\dot{X_{c}}}\right]} \dot{X_{b}}
\nonumber\\
&&\ \ \ \ \,    +
F_{X_{a} T} \ddot{T}+ \sum_{b=1}^{2}{F_{X_{a} X_{b}}\ddot{X_{b}}},\\
&&\!\!\!\!\!\!\!\!\!\! 
\dddot{F}_{X_{a}}=\frac{\partial 
\ddot{F}_{X_{a}}}{\partial 
{T}}\dot{T}+\sum_{b=1}^{2}{\frac{\partial \ddot{F}_{X_{a}}}{\partial {X_{b}}}\dot{X}_{b}},
\end{eqnarray} 
with   $F_{TT}=\partial^2F/\partial T^{2}$, $F_{TTT}=\partial^3F/\partial T^{3}$, 
$F_{X_{a} 
X_{b}}=\partial^2F/\partial X_{a}\partial X_{b}$ and $F_{X_{a} X_{b} 
X_{c}}=\partial^{3}F/\partial 
X_{a}\partial X_{b}\partial X_{c}$. 
Additionally, note that in the FRW geometry (\ref{weproudlyuse}), the torsion scalar 
(\ref{Tdefscalar}), as well as the functions $X_{1}$ and $X_{2}$, become
\begin{eqnarray}
\label{Tscalar2}
&& T=-6 H^2,\\
&& X_{1}=144 H^{2}\dot{H}^{2}\label{Xone},\\
&& X_{2}=-12 \left[\dot{H}\left(\dot{H}+3 H^{2}\right) + H \ddot{H}\right].
\label{Xtwo}
\end{eqnarray}
Finally, note that in the   Friedmann equations (\ref{Fr1}),(\ref{Fr2}) we have assumed 
that the matter energy-momentum tensor corresponds to a perfect fluid with 
energy density $\rho_m$ and pressure $p_m$.

The Friedmann equations (\ref{Fr1}),(\ref{Fr2}) can be re-written in the standard form 
\begin{eqnarray}
  3H^2&=&\rho_{DE}+\rho_m, \label{SFr1} \\
  -2\dot{H}& =&\rho_{m}+p_{m}+\rho_{DE}+p_{DE}, 
\label{SFr2}
\end{eqnarray}
where the   energy density and pressure of the effective dark energy sector are 
respectively
defined as
\begin{eqnarray}
\label{rhode}
&& 
\!\!\!\!\!\!\!
\rho_{DE}\equiv-\frac{F}{2}-6 H^2 F_{T}+3H^2
\nonumber\\
&&
\ \ \
-6F_{X_{2}}\dot{H}^2
-6 H \left(F_{X_{2}}+24 H^2F_{X_{1}}\right)\left(3 H \dot{H} +\ddot{H}\right) 
\nonumber\\
&&\ \ \ 
-144 H^3 
\dot{H}\dot{F}_{X_{1}}-6 H(3 H^2- \dot{H}) \dot{F}_{X_{2}} -6 
H^2\ddot{F}_{X_{2}},
\end{eqnarray}
\begin{eqnarray}\label{pde}
&&\,
p_{DE}\equiv 
-3{{H}^{2}}-2\dot{H}+2\Bigg\{
 {F}_{T} \dot{H}+ H \dot{F}_{T}\nonumber
\\
&&\ \ \ \ \ \ \ \ \ \     \,
+24 H \left[2 H \ddot{H} +3\left( 
\dot{H}+H^{2}\right)\dot{H} \right]\dot{F}_{X_{1}}
\nonumber
\\
&&\ \ \ \ \ \ \ \ \ \   \,
+12 H \dot{H} \dot{F}_{X_{2}}+24 
H^{2}\dot{H}\ddot{F}_{X_{1}}+\left(\dot{H} +3 
H^{2}\right)\ddot{F}_{X_{2}}\nonumber
\\
&&\ \ \ \ \ \ \ \ \ \    \,
+24 H^{2} {F}_{X_{1}} \dddot{H}+H\dddot{F}_{X_{2}}+24 {{F}_{X_{1}}} {{\dot{H}}^{2}}\left( 
12 {{H}^{2}}+\dot{H}\right)\nonumber \\
&&\ \ \ \ \ \ \ \ \ \   \,
+24 
H {F}_{X_
{1}}\left( 4 \dot{H} +
3 {{H}^{2}}\right)\ddot{H}\Bigg\}.
\end{eqnarray} 
As we mentioned earlier, since we have considered a matter sector minimally coupled to 
gravity and diffeomorphism invariant, the covariant derivative of the matter 
energy-momentum tensor is zero, which in the case of perfect fluid in FRW geometry leads 
to
\begin{equation}
\dot{\rho}_{m}+3H(\rho_{m}+p_{m})=0.
\label{EoSmat}
\end{equation}
Hence, in this case, the Friedmann equations (\ref{SFr1}),(\ref{SFr2}) imply 
 \begin{eqnarray}
\dot{\rho}_{DE}+3H(\rho_{DE}+p_{DE})=0.
\end{eqnarray}
In the following we will consider the matter fluid to be of barotropic nature, 
namely  $p_m=(\gamma-1)\rho_m$, with 
$w_{m}\equiv\gamma-1$ its equation-of-state parameter. Similarly, we  can define the 
effective dark energy equation-of-state parameter as
\begin{equation}
w_{DE}=\frac{p_{DE}}{\rho _{DE}}.
\label{wDE1}
\end{equation}
Lastly, concerning cosmological investigations it proves convenient to introduce the 
standard density 
parameters  $\Omega_{m}\equiv\frac{\rho_{m}}{3 H^2}$ and 
$\Omega_{DE}\equiv\frac{\rho_{DE}}{3 H^2}$, as well as the 
total equation-of-state parameter as 
\be
w_{tot}=\frac{p_{DE}+p_m}{\rho _{DE}+\rho _m},
\label{wtot}
\ee
which is immediately related to the deceleration parameter $q$ through 
\be
q=\frac{1}
{2}\left(1+3w_{tot}\right),
\label{deccelparam}
\ee
and hence acceleration occurs when $q<0$.

In summary, in the  cosmological scenario of modified gravity with higher-order torsional 
derivatives, one obtains an effective dark energy sector that comprises of these novel 
torsional terms. As we observe from the specific expressions, although TEGR coincides 
completely with general relativity at the level of equations, the corresponding 
modified scenario is different from its curvature analogue. This is a common feature of 
all torsional modified gravities, namely that they do not coincide with their curvature 
analogues, despite the fact that their starting theories are equivalent. Hence, since the 
present scenario is a novel class of gravitational modification, it is both interesting 
and necessary to investigate its cosmological applications. This is performed in the next 
section.

\section{Cosmological dynamics}
\label{CosmDynamics}

In the previous section we presented a torsional modified gravity based on the use of 
higher derivative terms, and we applied it in a cosmological framework. As we saw, 
in such a scenario we have obtained an effective dark energy sector which arises from the 
novel, higher-derivative torsional terms. In this section we are interested in 
investigating in detail the cosmological dynamics, using the powerful method of 
dynamical-system analysis \cite{Coley:2003mj}, which allows  to by-pass the 
complexities of the equations and reveal the global behavior of the system.

In order to perform the phase-space analysis of the cosmological scenario at hand, we 
have to introduce suitable dimensionless auxiliary variables that will bring the system 
of cosmological equations into its autonomous form 
\cite{Coley:2003mj,Copeland:1997et}.
 For a system of order 
$l$ we introduce the following dimensionless variables  \cite{Coley:2003mj}
\begin{eqnarray}
Z_{1}=H, \:\:\: Z_{2}=\frac{\dot{H}}{H^2},...,Z_{l+1}=\frac{\stackrel{l}{H}}{H^{l+1}},
\label{DynVariables}
\end{eqnarray} with $l=1,...,n$, the number of overdots. 
Then, the field equations can be rewritten in the form of an autonomous system 
\cite{Coley:2003mj}
\begin{eqnarray}\label{GeneralAuto}
&&\frac{dZ_{1}}{dN}={Z}_{1} {Z}_{2},\nonumber\\
&& \frac{dZ_{2}}{dN}={{Z}_{3}}-2 Z_{2}^2,\nonumber\\
&& \:\:\:\:\:\:\:\:\: \:\:\:\: \vdots \\
&& \frac{dZ_{l+1}}{dN}=Z_{l+2}-(l+1)Z_{2}Z_{l+1}\nonumber,
\end{eqnarray}
where we have introduced the e-folds number $N=\log{a}$. The system is 
truncated at 
the variable $Z_{l+2}$. Thus, the dimensionless variable 
$Z_{l+2}=Z_{l+2}(Z_{1},Z_{2},...,Z_{l+1})$ 
is calculated from the field equations. 

The critical points $(Z_{1}^{*},Z_{2}^{*},..,Z_{l+1}^{*})$ of the above dynamical system 
can be extracted by imposing the conditions 
$\frac{dZ_{1}}{dN}=\frac{dZ_{2}}{dN}=...=\frac{dZ_{l+1}}{dN}=0$. Observing 
(\ref{DynVariables}) and \eqref{GeneralAuto}  we can easily deduce that a de Sitter 
critical point 
is realized if
\begin{equation}
Z_{1}^{*}>0,\:\:\:\: Z_{2}^{*}=Z_{3}^{*}=...=Z_{l+2}^{*}=0, 
\label{deSitterSI}
\end{equation}
since in this case we immediately obtain  $a(t)\sim e^{H t}$, with $H=Z_{1}^{*}>0$ and  
$l=1,...,n$.
Similarly, a power-law form for the scale factor is realized if
\begin{equation}
Z_{1}^{*}=0,\:\:\:\: Z_{l+2}^{*}=l\,!(l+1)Z_{2}^{*(l+1)}, 
\label{PowerLaw}
\end{equation}
for   $l=1,...,n$,
in which case asymptotically we have  $a(t)\sim t^{p}$  with $p=-1/Z_
{2}^{*}$ (note that $Z_{l+1}$ for   $l=1,...,n$ can be non-zero although both the 
numerators and denominators in their definitions (\ref{DynVariables}) tend to zero). 

Finally, perturbing the system linearly around these critical points, and expressing the 
perturbations equations
in terms of a perturbation matrix, allows one to determine the type and stability of 
each  critical point by examining the eigenvalues of this matrix 
\cite{Coley:2003mj,Copeland:1997et}. 

In the following we apply this procedure to two 
specific $F(T,\left(\nabla{T}\right)^2,\Box {T})$ models.

\subsection{Model I: $F(T,\left(\nabla{T}\right)^2,\Box {T})=T+    
\frac{\alpha_{1}\left(\nabla{T}\right)^2}{T^2}+\alpha_{2} e^{\frac{\delta 
\left(\nabla{T}\right)^2}{
T^4}}$}
\label{mod11}

\def\tablename{Table}%
\begin{table*}[ht]
\centering
\begin{center}
\begin{tabular}{c c c c c c c c c c}\hline\hline
Name &  $\left\{Z_{1}^{*},Z_{2}^{*},Z_{3}^{*}\right\}$&   Existence  &\ \  $a(t)$  
& 
$\Omega_{DE}$ & $w_{DE}$ & $q$ &  Expansion  & Acceleration  &  
Stability 
\\\hline 
\\ 
$P_{1}$ & $\left\{0,-\frac{3 \gamma}{2},\frac{9 {{\gamma}^{2}}}{2}\right\}$ &Always & 
$t^{\frac{2}{3\gamma}}$    & $\frac{3 \left(4-\gamma\right)\gamma\alpha_{1}}{2}$ & 
$\gamma-1$& $-1+\frac{3\gamma}{2}$ &  Always &
$\gamma<\frac{2}{3}$     & 
Stable \\ \\ 
$P_{2-}$ & $\left\{0,Z^{*}_{2-}, 2 Z^{* 2}_{2-}\right\}$  &  $\ \alpha_{1}<0  $ or
$\alpha_{1}\geq 
\frac{1}{6}\ \ $ 
&$t^{ -\frac{1}{{{Z}_{2-}^{*}}} }
$   & 
$1$
 & $-1-\frac{2{{Z}_{2-}^{*}}}{3} $  
 & $-1- {Z}_{2-}^{*} $ & $\alpha_{1}\geq\frac{1}{6}$
 & No  
 & Saddle \\ \\ 
$P_{2 +}$ 
& $\left\{0,Z^{*}_{2+}, 2 Z^{* 2}_{2+}\right\}$  &  $\ \alpha_{1}<0  $ or
$\alpha_{1}\geq 
\frac{1}{6}\ \ $ 
&$t^{-\frac{1}{{{Z}_{2+}^{*}}}}$   & 
$1$
    & $-1-\frac{2{{Z}_{2+}^{*}}}{3}$ &
    $\ \ -1- {Z}_{2+}^{*}\ \ $ & Always&
$\alpha_{1}>\frac{3}{10}$
    & Stable    \\ 
    &&&&&&& &&  for  $\alpha_{1}>\frac{2}{3\gamma\left(4-\gamma\right)}$
    \\  
$P_{3}$ 
& $\left\{\sqrt{\frac{-\alpha_{2}}{6}},0, 0\right\}$  &  $\ \alpha_{2}<0$
& $e^{\sqrt{\frac{-\alpha_{2}}{6}}t}$   & 
$1$
    & $-1$ &
    $\ \ -1\ \ $ & Always&
Always
    & Stable    \\ 
    &&&&&&& &&  for $\delta<-\alpha_{1} \alpha_{2}$
    \\  
\hline\hline
\end{tabular}
\end{center}
\caption{
The physical critical points of the system   \eqref{Auto} of Model I: 
$F(T,\left(\nabla{T}\right)^2,\Box {T})=T+    
\frac{\alpha_{1}\left(\nabla{T}\right)^2}{T^2}+\alpha_{2} e^{\frac{\delta 
\left(\nabla{T}\right)^2}{
T^4}}$,  their 
existence and 
stability conditions, the asymptotic behavior of the scale factor $a(t)$  along with the 
conditions for 
expansion and acceleration, and the 
corresponding 
values of the dark energy density parameter $\Omega_{DE}$, of the dark energy 
equation-of-state parameter $w_{DE}$, and of the deceleration parameter $q$.   We have 
defined 
${{Z}_{2\pm}^{*}}=\frac{3\left[-\alpha_{1}\pm \sqrt{\alpha_{1} 
\left(\alpha_{1}-\frac{1}{6}\right)}\right]}{\alpha_{1}}$.} 
\label{Table1}
\end{table*}

Let us start our analysis by a simple scenario, in which the action does not depend on 
$X_{2}\equiv \Box {T}$ but only on  $X_{1}\equiv \left(\nabla{T}\right)^2$, i.e a 
scenario 
of the form
\begin{eqnarray}
&&
\!\!\!\!\!\!\!\!\!\!\!\!\!\!\!\!\!\!\!\!
F(T, X_{1}, X_{2})= T+\frac{\alpha_{1} X_{1}}{T^2}+\alpha_{2} e^{\frac{\delta 
X_{1}}{T^4}}
\nonumber\\
&&
\ \ \ \ \ \ \ \ =
T+   \frac{\alpha_{1} \left(\nabla{T}\right)^2}{T^2}+\alpha_{2} 
e^{\frac{\delta\left(\nabla{T}\right)^2}{T^4}},
\label{ModelI}
\end{eqnarray}
where $\alpha_{1}$, $\alpha_{2}$ and $\delta$ are constants.
As described above, we introduce the following three dimensionless variables
\begin{equation}
Z_{1}=H, \:\:\: Z_{2}=\frac{\dot{H}}{H^2}, \:\:\:\: Z_{3}=\frac{\ddot{H}}{H^3}.
\end{equation}
In terms of these variables, the torsion scalar from (\ref{Tscalar2}) and the function 
$X_1$ from (\ref{Xone}) become
\begin{eqnarray}
&& T=-6 Z_{1}^2,\nonumber\\
&& X_{1}=144 {{Z}_{1}^{6}} {{Z}_{2}^{2}}.
\end{eqnarray}
Hence, the system of the two Friedmann equations (\ref{Fr1}),(\ref{Fr2}) is written in 
its autonomous form as
\begin{eqnarray}
&&\frac{dZ_{1}}{dN}={Z}_{1} {Z}_{2},\nonumber\\
&& \frac{dZ_{2}}{dN}={{Z}_{3}}-2 {{{Z}_{2}}^{2}},\nonumber\\
&& \frac{dZ_{3}}{dN}={{Z}_{4}}-3 {{Z}_{2}} {{Z}_{3}},
\label{Auto}
\end{eqnarray} 
where the function $Z_{4}(Z_{1},Z_{2},Z_{3})$ is calculated from Eqs. \eqref{SFr1} and 
\eqref{SFr2} and is given in (\ref{ZModelIExam1}).
Additionally, in terms of the auxiliary variables, and using (\ref{rhode}),(\ref{pde}) 
and (\ref{wDE1}),(\ref{deccelparam}),  we can express the observables as 
 \begin{eqnarray}
&&
\!\!\!\!\!\!\!\!\!\!\!\!\!\!\!\! 
{{\Omega}_{\mathit{DE}}}=
\frac{\alpha_{2}}{486 
Z_{1}^6} e^{\frac{Z_{2}^2 \delta}{9 Z_{1}^2}}
\Big[
18 Z_{1}^2 (4 Z_{2}^2-Z_{3}-3 Z_{2})\delta
\nonumber\\
&& \ \ \ \  \ \ \ \ \ \ \ \ \ \ \ \  \  \ \,
+4 Z_{2}^2 (3 
Z_{2}^2-Z_{3})\delta^2
-81 Z_{1}^4\Big]
\nonumber\\
&&  \ \ 
+\frac{2  \alpha_{1} }{3} \left(3 
Z_{2}^2-2 Z_{3}-6 Z_{2}\right),
\label{Ommod1}
\end{eqnarray}
\begin{eqnarray}
&& 
\!\!\!\!\!\!\!\!\!\!\!\!\!\!\!\!\!\!\!\!\!
{{w}_{\mathit{DE}}}
=\gamma-1+\Big[
162 
Z_{1}^6 (
2 Z_{2}+3 \gamma)
 \Big]
\nonumber\\
&&
\times
\Big\{
\alpha_{2}\, e^{\frac{Z_{2}^2 \delta}{9 Z_{1}^2}}
\Big[
18 Z_{1}^2 (Z_{3}-4 Z_{2}^2+3 
Z_{2})\delta
\nonumber\\
&&
\ \ \ \ \ \ \ \ \ \ \ \ \ \ \ \,
+ 4 Z_{2}^2 (Z_{3}-3 Z_{2}^2)\delta^2
+
81 Z_{1}^4\Big]
\nonumber\\
&&\ \ \ \ 
+324 \alpha_{1} Z_{1}^6 (2 Z_{3}-3 Z_{2}^2+6 
Z_{2}) 
\Big\}^{-1},
\label{wdemod1}
\end{eqnarray}
and
\begin{equation}
q=-1- Z_2.
\label{qmod1}
\end{equation}

The scenario of Model I admits four physical critical points (i.e. real 
and corresponding to 
$0\leq\Omega_{DE}\leq1$), which are displayed in Table 
\ref{Table1} along with their existence conditions. In the same Table we include   
the asymptotic behavior of the scale factor $a(t)$ along with the conditions for 
expansion and acceleration, as well as the corresponding 
values of the dark energy density parameter $\Omega_{DE}$  calculated from  
(\ref{Ommod1}), of the dark energy 
equation-of-state parameter $w_{DE}$ from  
(\ref{wdemod1}),  and of the deceleration parameter $q$ from  
(\ref{qmod1}).  As we can see from the coordinates of the critical points $P_{1}$ and 
$P_{2\pm}$  
in Table \ref{Table1}, they satisfy the constraint \eqref{PowerLaw}, and thus they 
correspond to power-law solutions. On the other hand, the critical point $P_{3}$ 
satisfies 
the constraint \eqref{deSitterSI} and thus it is a de Sitter solution. Finally, in 
Table \ref{Table1} we include 
the stability conditions, arising from the investigation of Appendix \ref{App1}.

Point $P_1$ is stable and thus it can attract the universe at late 
times. It corresponds to an expanding universe in which the
dark-energy density parameter lies in the interval $0<\Omega_{DE}<1$, and therefore it 
could alleviate the coincidence problem since in this point  the dark energy and 
matter density parameters are of the same order of magnitude. However, for usual dust 
matter it cannot lead to acceleration, and hence this point is not favored by 
observations.
\begin{figure}[htbp]
\centering
\includegraphics[width=0.5\textwidth,trim=4 4 4 4,clip]{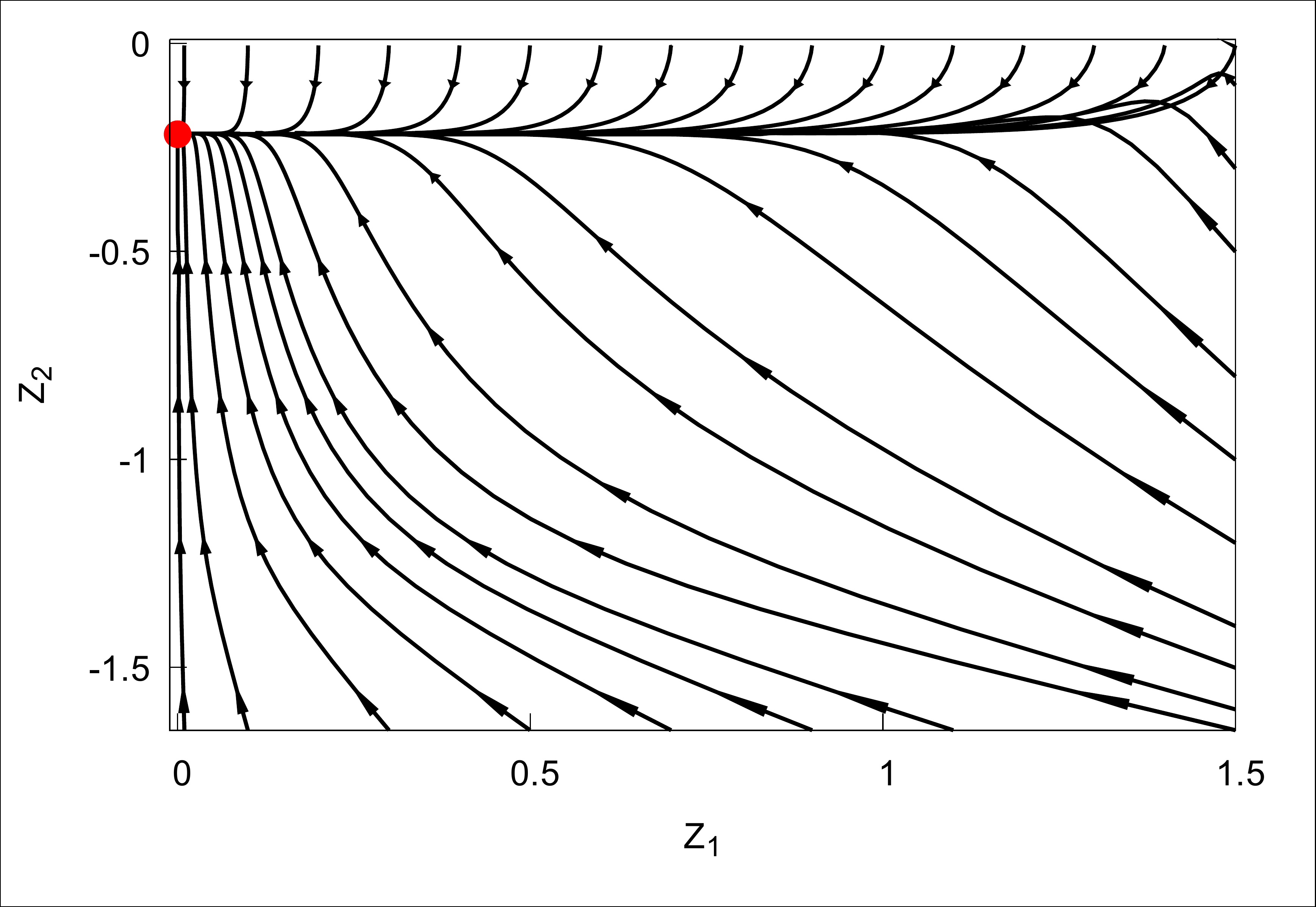}
\caption{\it{
The projection of the phase-space evolution on the  $Z_1-Z_2$
plane, for Model I: $F(T,\left(\nabla{T}\right)^2,\Box {T})=T+    
\frac{\alpha_{1}\left(\nabla{T}\right)^2}{T^2}+\alpha_{2} e^{\frac{\delta 
\left(\nabla{T}\right)^2}{
T^4}}$, for $\gamma=1$, $\alpha_{1}=1.2$ and $\alpha_{2}=0$,  in 
units where $8 \pi G=1$. The universe is attracted by the   
quintessence-like stable point $P_{2 +}$, marked by the red bullet.}}
	\label{Fig1}
\end{figure}

Point  $P_{2-}$ corresponds to a dark-energy dominated ($\Omega_{DE}=1$) universe, which 
can be expanding for some region of the parameter value $\alpha_{1}$, but it can never be
accelerating. However, it is a saddle point and therefore it cannot represent the 
late-time universe.

Point $P_{2+}$ corresponds to an expanding, dark-energy dominated universe, which can 
be accelerating for a large region of the model parameter $\alpha_{1}$, with the scale 
factor having a power-law form. Its corresponding dark energy equation-of-state parameter 
lies always in the quintessence regime ($w_{DE}\geq -1$), and it can acquire values very 
close to the 
observed ones for large model parameter $\alpha_{1}$ (for instance $w_{DE}\approx -0.98$ 
for $\alpha_{1}=10$ in units where $8 \pi G=1$). This fixed point can be stable for a 
large region of the model parameters. 

Point $P_{3}$ corresponds to an expanding, de Sitter solution, with $\Omega_{DE}=1$ and 
equation of state  $w_{DE}=w_{tot}=-1$. This fixed point is always accelerating and it is 
an 
attractor for a large region of the model parameters.

In summary, $P_{2+}$ and $P_{3}$ are the most important solutions in the scenario 
at hand, since they are both stable and possess observables in agreement with 
observations.

In order to present the above behavior in a more transparent way, we evolve   
the cosmological equations numerically and in Fig. \ref{Fig1} we depict the 
corresponding phase-space behavior projected on the $Z_1-Z_2$ plane, for given values of 
the model parameters  $\alpha_{1}$, $\alpha_{2}$ and $\delta$. As we can see, in 
this specific example the universe results in the  dark-energy dominated, accelerating,
quintessence-like stable point $P_{2 +}$. 

\begin{figure}[htbp]
\begin{center}
\includegraphics[width=0.5\textwidth,trim=4 4 4 4,clip]{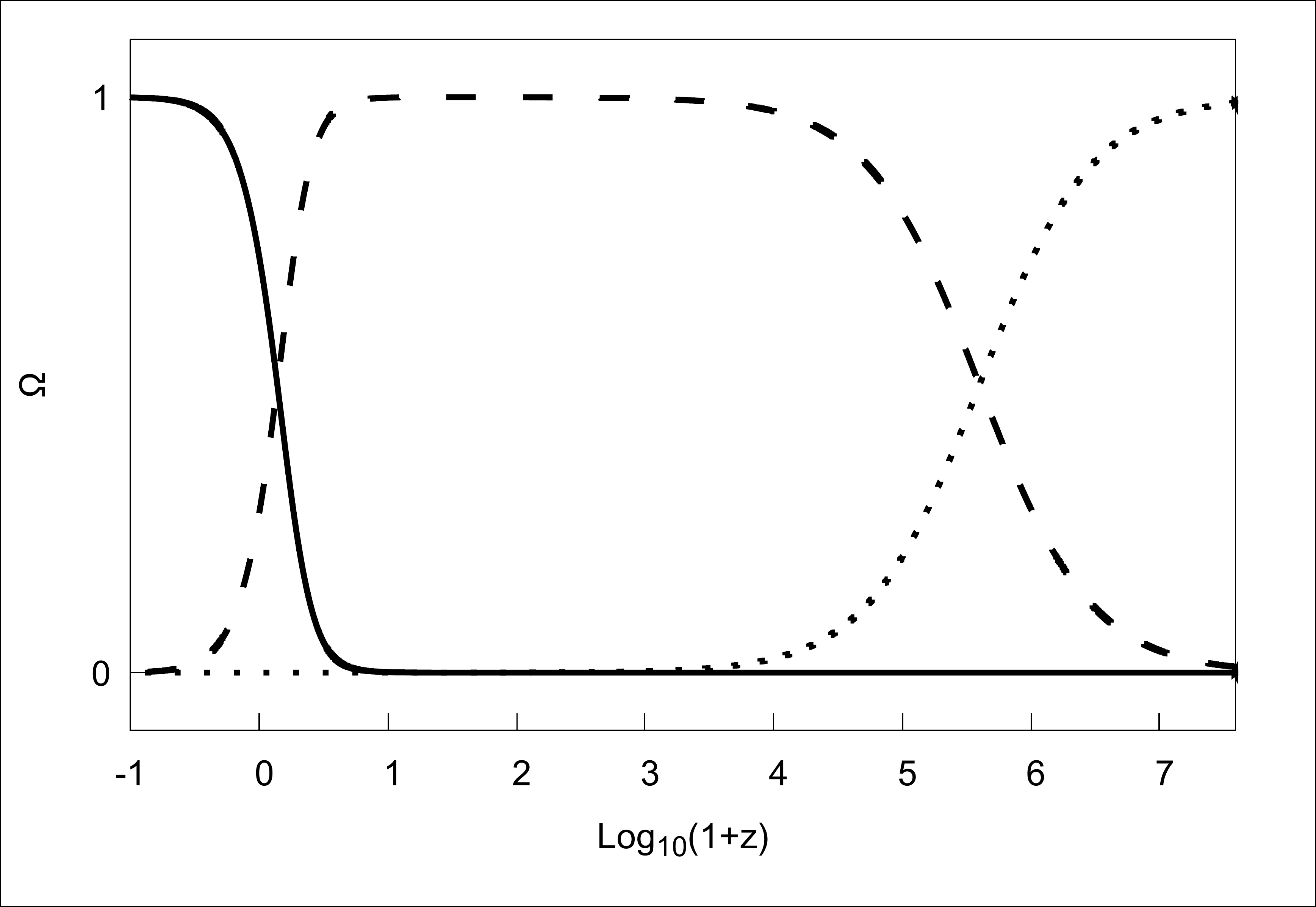}\\
  \includegraphics[width=0.5\textwidth,trim=4 4 4 4,clip]{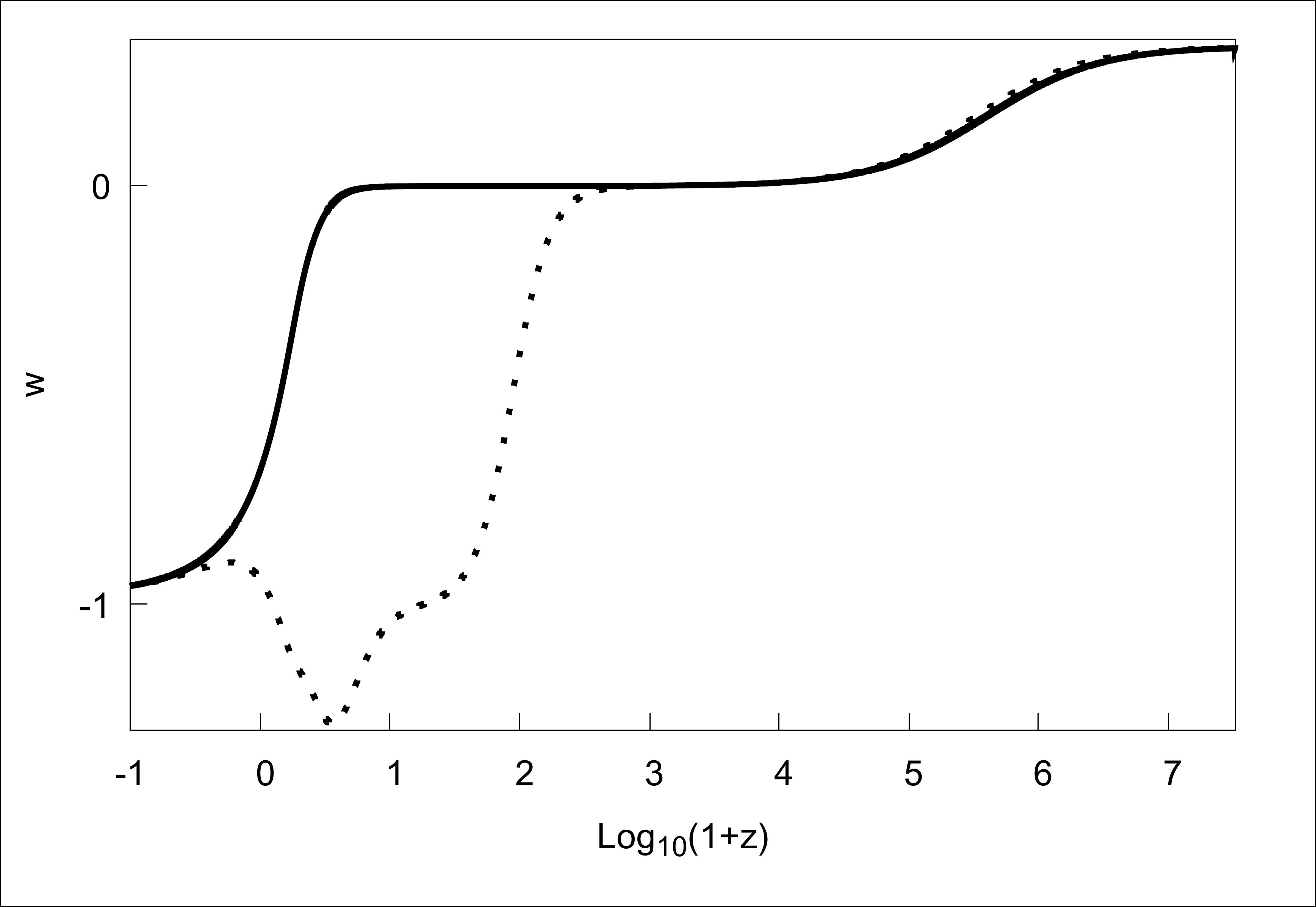}
\end{center}
\caption{\it{Evolution of various observables as a function of the redshift 
($z=\frac{a_0}{a}-1$ and for implicitly we set $a_0=1$), for
Model I: $F(T,\left(\nabla{T}\right)^2,\Box {T})=T+    
\frac{\alpha_{1}\left(\nabla{T}\right)^2}{T^2}+\alpha_{2} e^{\frac{\delta 
\left(\nabla{T}\right)^2}{T^4}}$, for $\alpha_{1}=3\times 10^{-7}$, $\alpha_{2}=-2\times 
10^{-20}$ and $\delta=-1\times 
10^{-19}$, in units where $8 \pi G=1$. In the upper graph we depict the evolution of the 
various density parameters, namely $\Omega_{DE}$ (solid line),  $\Omega_{m}$ (dashed 
line), and $\Omega_{r}$ (dotted line). In the lower graph we present the evolution of the 
dark-energy  (dotted line) and total (solid line) equation-of-state parameters.
The universe is attracted by the de Sitter fixed point $P_{3}$. For the numerics we have 
imposed $\Omega_{DE0}\approx 0.72$, $\Omega_{m0}\approx 0.28$,
$w_{DE0}\approx -0.94$ and $w_{tot0}\approx -0.68$ at present ($z=0$), in agreement with 
observations. }}
\label{Fig3}
\end{figure}

Apart from the correct late-time 
behavior, one should examine whether at early and intermediate times one can reconstruct 
the standard thermal history of the universe too. Hence, we numerically integrate the 
Friedmann equations (\ref{SFr1}),(\ref{SFr2}), including for completeness the radiation 
sector, and in the upper graph of Fig. \ref{Fig3}  we depict the evolution of the various 
density parameters as a function of the redshift ($z=\frac{a_0}{a}-1$ and for 
implicitly we set $a_0=1$). As we can see, we do obtain the successive sequence of 
radiation, matter, and dark energy epochs, as required, with the dark-energy fixed point 
being the de Sitter fixed point $P_{3}$. Additionally, in the lower graph of Fig. 
\ref{Fig3} we present the evolution of the dark-energy and total equation-of-state 
parameters, which is in agreement with the observed one. Note that for the parameter 
choices of the figure, $w_{DE}$ exhibits the phantom-divide crossing during the 
cosmological evolution, which is an advantage of the model.  

\subsection{Model II: $F(T,\left(\nabla{T}\right)^2,\Box {T})=T+\frac{\beta_{1}\Box 
{T}}{T}+\frac{\beta_{2}\left(\Box {T}\right)^2}{T^3}+\beta_{3} e^{\frac{\sigma \Box 
{T}}{T^3}}$}\label{mod2}

Let us now consider a scenario in which the action does not depend on 
 $X_{1}\equiv \left(\nabla{T}\right)^2$ but only on $X_{2}\equiv \Box {T}$, i.e a 
scenario of the form
\begin{eqnarray}
&&
\!\!\!\!\!\!
F(T, X_{1}, X_{2})= T+ \frac{\beta_1 X_{2}}{T}+\frac{\beta_2 X_{2}^2}{T^3}+\beta_{3} 
e^{\frac{\sigma X_{2}}{T^3}}
\nonumber\\
&& \ \ \ \ \ \ \ \ \ \ \ \ \ \ \   
=
T+\frac{\beta_1 \Box {T}}{T}+\frac{\beta_2 \left(\Box {T}\right)^2}{T^3}+\beta_{3} 
e^{\frac{\sigma \Box 
{T}}{T^3}},
\label{MII}
\end{eqnarray}
with $\beta_1$, $\beta_2$, $\beta_{3}$ and $\sigma$ the model parameters. We 
introduce the 
following five 
dimensionless variables
\begin{eqnarray}
&&
Z_{1}=H, \:\:\: Z_{2}=\frac{\dot{H}}{H^2}, \:\:\:\: Z_{3}=\frac{\ddot{H}}{H^3},\nonumber\\
&&
Z_{4}=\frac{\dddot{H}}{H^4},\:\:\:\: {{Z}_{5}}=\frac{\ddddot{H}}{H^5}.
\end{eqnarray}

\def\tablename{Table}%
\begin{table*}[ht!]
\centering
\begin{center}
\begin{tabular}{c c c c c c c c c c}
\hline\hline
Name &  $\left\{Z_{1}^{*},Z_{2}^{*},Z_{3}^{*},Z_{4}^{*},Z_{5}^{*}\right\}$ & Existence   
 &  $a(t)$ &    $\Omega_{DE}$ & $w_{DE}$ & $q$  &  Expansion &  Acceleration &  
Stability 
\\\hline \\ 
$Q_{1}$
& $\left\{0,-\frac{3 \gamma}{2},\frac{9 {{\gamma}^{2}}}{2},-\frac{81 {{\gamma}^{3}}}{4}, 
\frac{243 {
{\gamma}^{4}}}{2} 
\right\}$    &Always   & $t^{\frac{2}{3\gamma}}$    & 
$s(\gamma) \beta_1$ & $\gamma-1$  &  $-1+\frac{3\gamma}{2}$ & Always & 
$\gamma<\frac{2}{3}$  & Saddle
\\ \\ 
$Q_{2 -}$
& $\left\{0,A_-, 2A_-^2, 6 A_-^3, 24 A_-^4\right\}$ & 
$\beta_1\leq-0.71$  & 
$t^{ -\frac{1}{A_-} }$ 
  & 
$1$ & $-1-\frac{2A_-}{3}$ &    $-1-A_- $  & Always &No & 
Saddle 
\\ 
&& or $\beta_1>0$    &&&&&&&  
 \\ \\
$Q_{2 +}$
& $\left\{0,A_+, 2 A_+^2, 6 A_+^3, 24 A_+^4 \right\}$ & 
$\beta_1\leq-0.71$  & 
$t^{ -\frac{1}{A_+} }$ 
  & 
$1$ & $-1-\frac{2A_+}{3}$ &    $-1-A_+$  &No & Always & 
Saddle
\\ 
&&  or $\beta_1>0$   &&&&&&&  
 \\ \\
 $Q_{3-}$
 & $\left\{0,B_-, 2 B_-^2, 6 B_-^3, 24 B_-^4 \right\}$ &      $\beta_1\leq-0.
71$         &  
$t^{-\frac{1}{B_-}}$ &  
$1$ & $-1-\frac{2B_-}{3}$ & $-1- B_-$  & Always &No& Saddle
 \\ 
&&  or     $\beta_1>0.29$
  &&&&&&&  
 \\ \\
$Q_{3+}$ 
& $\left\{0,B_+, 2 B_+^2, 6 B_+^3, 24 
B_+^4 \right\}$ 
& $\beta_1\leq-0.
71$ 
 &  
$\ t^{-\frac{1}{B_+}}\  $ &  
$1$ & $\ -1-\frac{2B_+}{3}\ $ & $-1- B_+ $  &    $\beta_1>0.29$        &$\beta_1>0.3$ & 
Stable \\ 
&&  or       $\beta_1>0.29$
 &&&&&&& for $\beta_1>0.3$ 
 \\ \\
$Q_{4}$ 
& $\left\{\sqrt{\frac{-\beta_{3}}{6}},0, 0, 0, 0\right\}$ 
& $\beta_{3}<0$ 
 &  
$\ e^{\sqrt{\frac{-\beta_{3}}{6}} t}\  $ &  
$1$ & $\ -1$ & $-1$  &   Always      & Always& 
Stable
\\
\hline\hline
\end{tabular}
\end{center}
\caption{
The physical critical points of the system   \eqref{AutoModelII} of Model II:  
$F(T,\left(\nabla{T}\right)^2,\Box {T})=T+\frac{\beta_1 \Box {T}}{T}+\frac{\beta_2 
\left(\Box {T}\right)^2}{T^3}+\beta_{3} e^{\frac{\sigma \Box 
{T}}{T^3}}$ for $\beta_2=\frac{7 
\beta_1}{34}$, their existence and 
stability conditions, the asymptotic behavior of the scale factor $a(t)$  along with the 
conditions for 
expansion and acceleration, and the 
corresponding 
values of the dark energy density parameter $\Omega_{DE}$, of the dark energy 
equation-of-state parameter $w_{DE}$, and of the deceleration parameter $q$.   We have 
defined 
$A_\pm=\frac{-17\pm\sqrt{625+28\sqrt{6\left(\frac{17}{\beta_1}+24\right)}}}{14}$, 
\quad 
$B_\pm=\frac{-17 \pm \sqrt{625-28 
\sqrt{6\left(\frac{17}{\beta_1}+24\right)}}}{14}$, and $s(\gamma)=\frac{3\gamma\left( 
2+3\gamma\right)\left( 7\gamma-16\right)\left( 21\gamma-34\right)}{544}$.  
} 
\label{Table2}
\end{table*}

In terms of these variables, the torsion scalar from (\ref{Tscalar2}) and the function 
$X_2$ from (\ref{Xtwo}) become
\begin{eqnarray}
&& T   =-6 Z_{1}^2,\nonumber\\
&&  X_{2}=-12 {{Z}_{1}^{4}}\left[ Z_{3}+{{Z}_{2}}\left(3+{{Z}_{2}}\right)\right].
\end{eqnarray}
Hence, the system of the two Friedmann equations (\ref{Fr1}),(\ref{Fr2}) is written in 
its autonomous form as
{\small{\begin{eqnarray}
&&\frac{dZ_{1}}{dN}={Z}_{1} {Z}_{2},\nonumber\\
&& \frac{dZ_{2}}{dN}={{Z}_{3}}-2 {{{Z}_{2}}^{2}},\nonumber
\\
&& \frac{dZ_{3}}{dN}={{Z}_{4}}-3 {{Z}_{2}} {{Z}_{3}},\nonumber\\
&&\frac{dZ_{4}}{dN}={{Z}_{5}}-4 {{Z}_{2}} {{Z}_{4}},\nonumber\\
&& \frac{d Z_{5}}{dN}={{Z}_{6}}-5 {{Z}_{2}} {{Z}_{5}}.
\label{Auto2}
\end{eqnarray}}}
The function $Z_{6}(Z_{1},...,Z_{5})$ is calculated from Eqs. \eqref{SFr1} and 
\eqref{SFr2}, and is given in (\ref{ZModelIExam2}).
Moreover, in terms of the auxiliary variables, and using (\ref{rhode}),(\ref{pde}) 
and (\ref{wDE1}),(\ref{deccelparam}),  we express the observables  $\Omega_{DE}$, 
$w_{DE}$ and $q$ respectively as 
{\small{\begin{eqnarray}
&&\!
{{\Omega}_{\mathit{DE}}}=
\frac{\beta_{3}}{34992{{Z}_{1}^{8}}}
e^{\frac{\left[Z_{3}+Z_{2}\left(Z_{2}
+3\right)\right]\sigma}{18 Z_{1}^2}}
\Big\{\!
-5832 Z_{1}^6
\nonumber\\
&& \ \  \ \  \ \  \ \  \ \  \ \ \
-324 Z_{1}^4 (11 Z_{3}-43 Z_{2}^2+33 
Z_{2})\sigma 
\nonumber\\
&&  \ \  \ \  \ \  \ \  \ \  \ \ \
+
18 Z_{1}^2
[Z_{5}+3 (13 Z_{2}^2-28 Z_{2}+3) Z_{3}-3 Z_{3}^2
\nonumber\\
&& 
\ \  \ \  \ \  \ \  \ \  \ \ \ \ \ \
-3 (7 Z_{2}-2) Z_{4}+6 Z_{2}^2 (20 Z_{2}^2+33 Z_{2}-6)]\sigma^2
\nonumber\\
&&  \ \  \ \  \ \  \ \  \ \  \ \ \
+[Z_{4}+3 (1-Z_{2}) Z_{3}-6 Z_{2}^3-12 Z_{2}^2]^2 \sigma^3
\Big\}
\nonumber\\
&& \ \ \ \ \ \ \ \ \ \, 
-
\frac{1}{9}
\Big\{
[ 2 {{Z}_{5}}-2 (2{{Z}_{2}^{2}}+63 
{{Z}_{2}}-9){{
Z}_{3}}-11 {{Z}_{3}^{2}}
\nonumber
\\ 
&& \ \  \ \  \ \  \ \  \ \  \ \ \ \ \ \ \ 
-6 ( 3 {{Z}_{2}}-2){{Z}_{4}}
+13 {{Z}_{2}^{2}}( 7 {{Z}_{2}^{2}}+6 
{{Z}_{2}}-9)]
\beta_{2}
\nonumber\\
&& 
\ \  \ \  \ \  \ \  \ \  \ \ \ \ \ 
+
6\Big(2{{Z}_{3}}-3 {{Z}_{2}^{2}}+6 {{Z}_{2}}\Big)\beta_{1}
\Big\},
\label{Ommod2}
\end{eqnarray}}}
\begin{eqnarray}
&&\!\!\!\!
{{w}_{\mathit{DE}}}=
\gamma-1
-11664 {Z}_{1}^{8}\Big( 2 {{Z}_{2}}+3\gamma\Big)
\nonumber\\
&& \ \  \ \  \ \ \
\times
\Bigg\{
\Big\{[Z_{4}+3 (1-Z_{2}) Z_{3}-6 Z_{2}^3-12 Z_{2}^2]^2 
\sigma^3
\nonumber\\
&&  \ \  \ \   \ \  \ \  \ \ \ \ \
+
18 Z_{1}^2 [Z_{5}\!+\!(39 Z_{2}^2\!-\!84 Z_{2}+9) Z_{3}\!-\!
3 Z_{3}^2\!-\!36 Z_{2}^2
\nonumber\\ 
&& \ \  \ \   \ \  \ \  \ \ \ \ \ \ \ \ \ \ \   \ \ \ \ 
\!+\!3 (2-7 Z_{2}) Z_{4}
\!+\!120 Z_{2}^4\!+\!198 Z_{2}^3]\sigma^2
\nonumber\\
&& 
 \ \  \ \   \ \  \ \  \ \ \ \ \
 -
324 Z_{1}^4 (11 Z_{3}-43 Z_{2}^2+33 Z_{2})\sigma
\nonumber\\
&&
 \ \  \ \   \ \  \ \  \ \ \ \ \
 -5832 Z_{1}^6
\Big\}\beta_{3}\,
e^{\frac{\left[Z_{3}+Z_{2}\left(Z_{2}+3\right)\right]\sigma}{18 Z_{1}^2}}
\nonumber\\
&& \  \ \   \ \  \ \  \  \
-
3888 Z_{1}^8 
\Big\{[2 Z_{5}-11 Z_{3}^2
-
2 (2 Z_{2}^2+63 Z_{2}-9) Z_{3}
\nonumber\\
&&
 \  \ \   \ \  \ \  \  \  \  \ \   \ \  \ \   -6 (3 Z_{2}-2) Z_{4}
+13 Z_{2}^2 (7 Z_{2}^2+6 Z_{2}-9)] \beta_{2}
\nonumber\\
&&
 \  \ \   \ \  \ \  \  \  \  \ \   \ \  \ \  
+
6 (2 Z_{3}-3 Z_{2}^2+6 Z_{2})\beta_{1}\Big\}
\Bigg\}^{-1},
\label{wdemod2}
\end{eqnarray}
and
\begin{equation}
q=-1- Z_2.
\label{qmod2}
\end{equation}

\begin{figure}[htbp]
\centering
\includegraphics[width=0.5\textwidth,trim=4 4 4 4,clip]{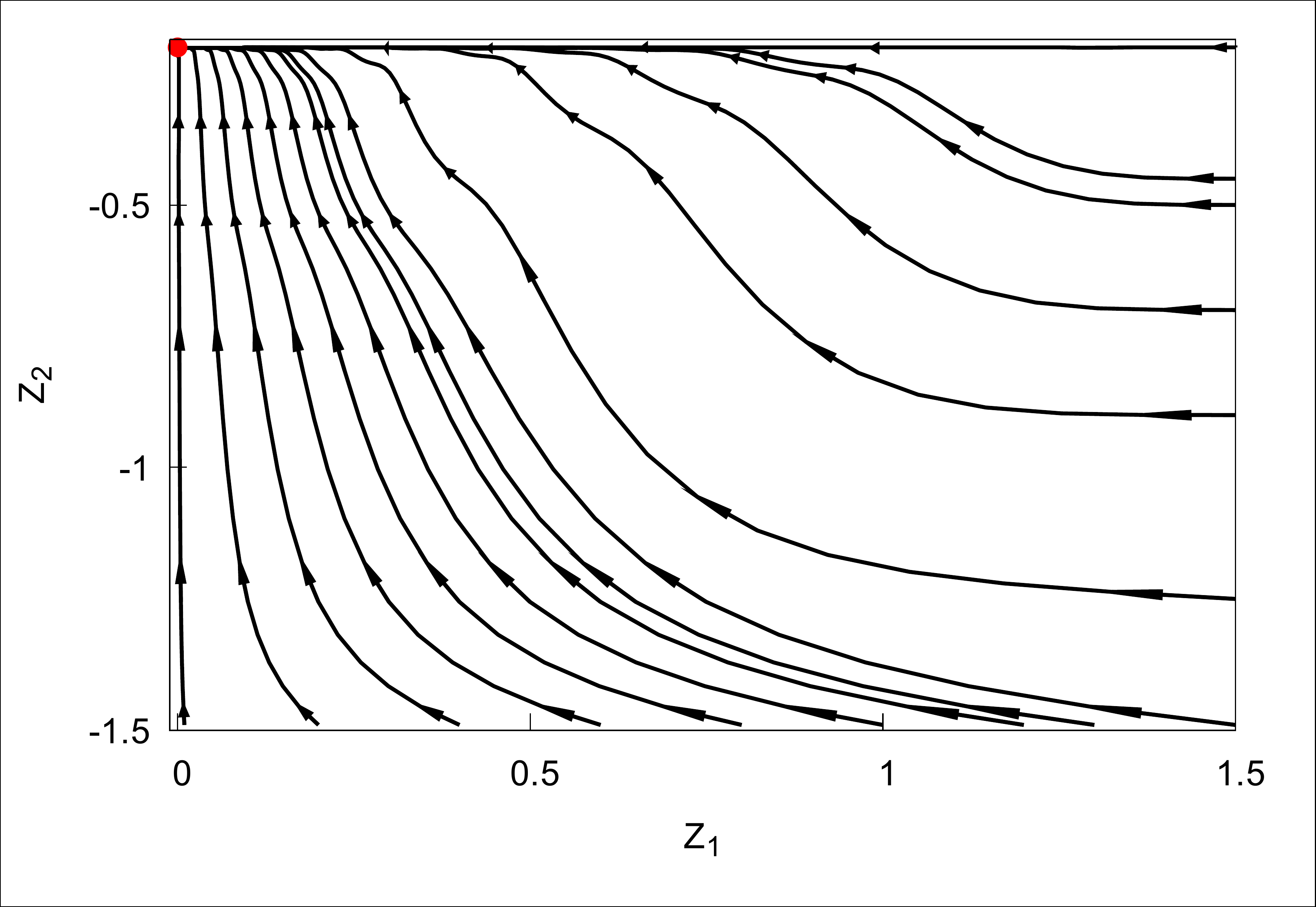}
\caption{\it{The projection of the phase-space evolution on the  $Z_1-Z_2$
plane, for Model II: $F(T,\left(\nabla{T}\right)^2,\Box {T})=T+\frac{\beta_1\Box 
{T}}{T}+\frac{\beta_2 \left(\Box 
{T}\right)^2}{T^3}+\beta_{3} e^{\frac{\sigma \Box {T}}{T^3}}$, for $\gamma=1$, 
$\beta_1=1.2$, 
$\beta_2=\frac{7\beta_1}{34}\approx 0.25$ and $\beta_{3}=0$, in 
units where $8 \pi G=1$. The universe is attracted by the   
quintessence-like stable point $Q_{3 +}$, marked by the red bullet.}}
	\label{Fig2}
\end{figure}

The system (\ref{Auto2}) cannot be analytically handled in the case of general 
$\beta_1$ and $\beta_2$. Hence, in order to analytically extract its critical points we 
need 
to assume a relation between these two coupling parameters. Without loss of generality, 
and in order to simplify the expressions, we consider $\beta_2=\frac{7\beta_1}{34}$. In 
this case, the scenario of Model II admits six physical critical points (i.e. real and 
corresponding to  $0\leq\Omega_{DE}\leq1$), which are displayed in Table \ref{Table2} 
along with their existence conditions. In the same Table we include the asymptotic 
behavior of the scale factor $a(t)$ along with the conditions for 
expansion and acceleration, as well as the corresponding values of the dark 
energy density parameter $\Omega_{DE}$  calculated from (\ref{Ommod2}), of the dark 
energy equation-of-state parameter $w_{DE}$ from (\ref{wdemod2}), and of the deceleration 
parameter $q$ from (\ref{qmod2}). As we can see from the coordinates of the critical 
points $Q_{1}$, $Q_{2\pm}$ and $Q_{3\pm}$ in Table \ref{Table2}, they satisfy the 
constraint \eqref{PowerLaw}, and thus they correspond to power-law solutions. On the 
other 
hand, the critical point $Q_{4}$ satisfies the constraint \eqref{deSitterSI}, and thus it 
corresponds to a de Sitter solution. Finally, we include the stability 
conditions, arising from the investigation of Appendix \ref{App2}.

Point $Q_1$ corresponds to an expanding universe in which the dark-energy density 
parameter lies in the interval $0<\Omega_{DE}<1$, and therefore it could alleviate the 
coincidence problem, however for usual dust matter it cannot lead to acceleration. It
is a saddle one, and hence it cannot attract the universe at late times, nevertheless it 
could be the state of the universe for large intermediate-time intervals, describing the 
matter era.
 
Points $Q_{2-}$ and $Q_{3-}$ correspond to a dark-energy dominated ($\Omega_{DE}=1$) 
expanding universe, which however is always decelerating, and thus not favored by 
observations. Both points are saddle, and therefore they cannot be the 
stable late-time solutions of the universe. Additionally, point $Q_{2+}$ corresponds to a 
contracting universe, and since it is saddle it cannot attract the universe at late times.

Point $Q_{3+}$ corresponds to a dark-energy dominated universe, which can be expanding 
and accelerating for a large region of the model parameter $\beta_1$, with the scale 
factor having a power-law form. Its corresponding dark energy equation-of-state parameter 
lies always in the quintessence regime, and it can acquire values very close to the 
observed ones for large model parameter $\beta_1$ (for instance $w_{DE}\approx -0.98$ for 
$\beta_1=10$ in units where $8 \pi G=1$). This point can be stable for a large region of 
the model parameters, in particular for the same range that it is accelerating.

Point $Q_{4}$ corresponds to a de Sitter solution with $\Omega_{DE}=1$ and equation of 
state $w_{DE}=w_{tot}=-1$, it is always accelerating, and it can be stable for a large 
region of the model parameters.

In summary, $Q_{3+}$ and $Q_{4}$  are the most important solutions in the scenario 
at hand, since they are both stable and possess observables in agreement with 
observations.

In order to present the above behavior   more transparently, we evolve   
the cosmological system numerically and in Fig. \ref{Fig2} we present the 
corresponding phase-space behavior projected on the $Z_1-Z_2$ plane. As we observe, in 
this specific example the universe results in the dark-energy dominated, accelerating,
quintessence-like stable point  $Q_{3+}$. 

Although the present Model II can correctly describe the universe at late times, it 
cannot provide a very satisfactory behavior at intermediate times, since the 
exponential term in the Lagrangian \eqref{MII} cannot remain small for sufficiently large 
times in order for the standard matter epoch to be reproduced, but still increase at late 
times in order to drive acceleration (the exponent  $\Box{T}/T^3$ of the exponential 
oscillates around zero). Hence, the obtained matter era has smaller duration than the 
observed one. One may bypass this problem by adding higher order terms in the exponent, 
such as the quadratic quotient $\left(\Box{T}/T^3\right)^2$, which can regularize 
the oscillations and thus keep the effective dark energy sector very small 
during a sufficient time in order to reproduce successfully the matter era. However, such 
a construction would result to a more complicated model, with more complicated 
phase-space behavior, whose detailed investigation, although necessary and interesting,  
lies beyond the scope of the present work.

\section{Discussions and final Remarks}
\label{Conclusions}

In the present work we  constructed classes of modified gravitational theories using 
higher-derivative torsional terms. In particular, since we know that in curvature gravity 
one may construct theories with higher-order derivatives in the extended action, which 
could be justified as arising from higher-loop corrections in the high-curvature regime, 
one could in principle follow the same direction in torsion formulation of gravity and 
consider similar corrections to the simple action of Teleparallel Equivalent of general 
relativity (TEGR). As it is a common feature of all torsional modified gravities, 
although TEGR coincides completely with general relativity at the level of equations, the 
corresponding modified scenario with higher-derivative torsion terms, such as 
$\left(\nabla T\right)^2$ and $\Box{T}$, is different from its curvature analogue, 
i.e. it is  a novel class of gravitational modification. Hence,  it is both interesting 
and necessary to investigate its cosmological applications.

Extracting the general cosmological equations, we saw that we obtained an effective dark 
energy sector that comprises of the novel torsional contributions. Similarly to the 
curvature analogue models, for suitable constructions these novel terms can have a 
significant contribution, although they arise from higher-order derivatives. We then  
used the powerful method of dynamical system analysis in order to bypass the complexities 
of the equations and obtain information about the global, asymptotic behavior of the 
universe. In particular, we extracted the stable critical points of the scenario, which 
can thus be the late-time state of the universe, calculating moreover the corresponding 
observables, such as the various density parameters and the deceleration and 
dark-energy equation-of-state parameters, as 
well as the asymptotic form of the scale factor. We examined two specific scenarios, one 
based on $\left(\nabla T\right)^2$  and one based on $\Box{T}$ terms.

In the first Model we found that for a wide range of the model parameters the universe 
can result in a dark-energy dominated, accelerating universe, where the dark-energy 
equation-of-state parameter lies in the quintessence regime. Additionally, apart from the 
correct late-time behavior, the model can describe the thermal history of the universe, 
i.e. the successive sequence of radiation, matter and dark energy epochs, which is a 
necessary requirement for any realistic scenario. Moreover, during the evolution the 
dark-energy equation-of-state parameter may exhibit the phantom-divide crossing, which is 
an additional advantage. Similarly, in the second Model, although the global behavior is 
richer, we also found that the universe will be led to a dark-energy dominated, 
accelerating, quintessence-like state, for a wide region of the parameter space. In both 
Models the scale factor behaves asymptotically either as a power law or as an exponential 
law, while for large parameter regions the exact value of the dark-energy 
equation-of-state parameter can be in great agreement with observations. We mention that 
the above behavior has been obtained without the use of an explicit cosmological 
constant, and it is a pure results of the novel higher-derivative torsion terms.

In summary, as we can see, modified gravity with higher-derivative torsion terms can be 
very 
efficient in describing the evolution of the universe at the background level. However, 
before considering it as a successful candidate for the description of nature it is 
necessary to perform a detailed investigation of its perturbations, since perturbative  
instabilities may always arise (for instance this is the case in  
the initial version of Ho\v{r}ava-Lifshitz gravity
\cite{Bogdanos:2009uj}, in the initial version of de Rham-Gabadadze-Tolley massive 
gravity \cite{DeFelice:2012mx}, etc). Although such a detailed and complete analysis of
the cosmological perturbations is necessary, its various complications and 
lengthy calculations make it more convenient to be examined in a separate project 
\cite{usfuture}. Nevertheless, for the moment we would like to mention that in the case 
of simple $f(T)$ gravity, the perturbations of which were examined in detail  
\cite{Chen:2010va}, one may obtain instabilities, but there are many classes of $f(T)$ 
ansantzes and/or parameter-space regimes, where the perturbations are well-behaved. This 
feature is a good indication that we could  expect to find a similar behavior in
modified gravity with higher torsion derivatives too, although we need to indeed 
verify this under a thorough perturbation analysis.

\begin{acknowledgments}
The authors would like to thank A. A. Deriglazov, W.~G.~Ram\'irez,  M. Kr\v{s}\v{s}\'ak 
and F. ~S.
~N.~Lobo 
for useful comments.
G.O. would like to thank CAPES (Programm PNPD) for financial support. This work was 
partially  supported 
by the JSPS KAKENHI Grant Number JP 25800136 and the research-funds 
given by  Fukushima University (K. Bamba).
This article is also based upon work from COST action CA15117 (CANTATA), supported by 
COST (European Cooperation in Science and Technology).

\end{acknowledgments}

\appendix

 \section{Stability analysis of Model I: $F(T,\left(\nabla{T}\right)^2,\Box 
{T})=T+\frac{\alpha_{1}   
\left(\nabla{T}\right)^2}{T^2}+\alpha_{2} e^{\frac{\delta 
\left(\nabla{T}\right)^2}{T^4}}$} 
\label{App1}

In this Appendix we investigate the stability  of Model I: 
$F(T,\left(\nabla{T}\right)^2,\Box {T})=T+\frac{\alpha_{1}  
\left(\nabla{T}\right)^2}{T^2}+\alpha_{2} e^{\frac{\delta 
\left(\nabla{T}\right)^2}{T^4}}$ 
of subsection \ref{mod11}. 
The autonomous form of the system is  given in (\ref{Auto}), namely
\begin{eqnarray}
&&\frac{dZ_{1}}{dN}={Z}_{1} {Z}_{2},\nonumber\\
&& \frac{dZ_{2}}{dN}={{Z}_{3}}-2 {{{Z}_{2}}^{2}},\nonumber\\
&& \frac{dZ_{3}}{dN}={{Z}_{4}}-3 {{Z}_{2}} {{Z}_{3}},
\label{Autoapp}
\end{eqnarray} 
where the function $Z_{4}(Z_{1},Z_{2},Z_{3})$ is calculated from Eqs. \eqref{SFr1} and 
\eqref{SFr2} and it reads 
\bea
&& Z_{4}=
-\Bigg\{
\Bigg\{54 Z_{1}^2 Z_{2}
\Big\{Z_{3}^2
+[(\gamma+1) Z_{2}-10 
Z_{2}^2] Z_{3}
\nonumber\\
&&\ \ \ \ \ \ \ \ \ \ \ \ \ \ \ \ \ \ \ \ \ \ \ \ \ \  \,
+16 Z_{2}^4-3(1+\gamma) Z_{2}^3\Big\} 
\delta^2
\nonumber\\
&&
-486 \delta Z_{1}^4 
\Big[(4 Z_{2}-\gamma-1) 
Z_{3}
-7 Z_{2}^3+4(1+\gamma) Z_{2}^2-3 \gamma Z_{2}\Big]
\nonumber\\
&& 
+8 Z_{2}^3 (Z_{3}-3 Z_{2}^2)^2 
\delta^3+   2187 \gamma Z_{1}^6\Bigg\} \alpha_{2}
e^{\frac{Z_{2}^2 \delta}{9 Z_{1}^2}}   
\nonumber\\
&&
-1458 Z_{1}^8  \alpha_{1}
\Big\{2 [4 Z_{2}-3(1+ \gamma)]Z_{3}-6 Z_{2}^3 +
9 \gamma Z_{2}^2-18 \gamma Z_{2}\Big\} 
\nonumber\\
&&
+2187 Z_{1}^8 (2 Z_{2}+3 
\gamma)\Bigg\}
\nonumber\\
&&
\times
\left\{9 Z_{1}^2 
\left[\alpha_{2} \delta (2 Z_{2}^2 \delta+9 Z_{1}^2) e^{\frac{Z_{2}^2 \delta}{9 
Z_{1}^2}}+324 Z_{1}
^6 \alpha_{1}\right]\right\}^{-1}.
\label{ZModelIExam1}
\eea
The autonomous system \eqref{Autoapp} 
admits  four physical critical points (i.e. real and corresponding to 
 $0\leq\Omega_{DE}\leq1$), which are displayed in Table 
\ref{Table1} along with their existence conditions. 

In order to examine the stability of these critical points, we perform linear 
perturbations around them as $Z_{i}=Z_{i}^{*}+\delta{Z_{i}}$, and thus we extract the 
perturbation equations as  $ \textbf{U}'={\mathcal{M}}\cdot
\textbf{U}$, where $\textbf{U}$ is the
column vector of the perturbations $\delta{Z_{i}}$, and  $\mathcal{M}$ is the  $3\times3$ 
matrix that contains the coefficients of the
perturbation equations. The 
non-zero components of    $\mathcal{M}$   read
\begin{eqnarray}
&& \mathcal{M}_{11}=Z_{2},
\nonumber
\\
&& \mathcal{M}_{12}=Z_{1},\nonumber
\\
&& \mathcal{M}_{22}=-4 {{Z}_{2}},\nonumber
\\
&& \mathcal{M}_{23}=1,\nonumber
\\
&& \mathcal{M}_{31}=\frac{\partial Z_{4}}{\partial Z_{1}},\nonumber
\\
&& \mathcal{M}_{32}=\frac{\partial {Z}_{4}}{\partial Z_{2}}-3 {{Z}_{3}},\nonumber
\\
&& \mathcal{M}_{33}=\frac{\partial {{Z}_{4}}}{\partial {{Z}_{3}}}-3 {{Z}_{2}}.
\end{eqnarray}
Hence, as usual, the eigenvalues of $\mathcal{M}$  determine the type and
stability of the specific critical point. In particular, if the eigenvalues have
negative real parts then the critical point is stable, if they have positive real parts 
then the   critical point is unstable, and if they have real parts of different sign then 
the 
critical point is a saddle one.

For the fixed point $P_{1}$ the three eigenvalues $\mu_i$ write as
\bea
&&
\!\!\!\!\!\!\!\!\!\!\!\!\!\!\!
\mu_{1}=-\frac{3\gamma}{2}, \nonumber \\
&&
\!\!\!\!\!\!\!\!\!\!\!\!\!\!\! \mu_{2,3}=\frac{1}{4 
\alpha_{1}}\Big\{-3\left(2-\gamma\right) 
\alpha_{1}
\nonumber\\
&&\ 
\pm\Big[9 {{\left( 
2-\gamma\right) 
}^{2}}{{\alpha_{1}}^{2}}- 12\alpha_{1}\left( 2-3\left( 4-\gamma\right) 
\gamma\alpha_{1}\right)\Big]\Big\}.
\eea 
From the corresponding value of $\Omega_{DE}=\frac{3 
\left(4-\gamma\right)\gamma\alpha_{1}}{2}$ depicted in Table \ref{Table1}, we deduce that 
 the physical condition $0<\Omega_{DE}<1$ requires  $0<\alpha_{1}<\frac{2}{3 
\left(4-\gamma\right)\gamma}$. Hence, for this region $P_{1}$ is always stable. In 
particular, for  $\frac{8}{3\left[ 3\left(4-\gamma\right) \gamma+4\right] 
}<\alpha_{1}<\frac{2}{3 \left(4-\gamma\right)\gamma}$ it is a 
stable node, whereas  
for $0<\alpha_{1}<\frac{8}{3\left[ 3\left( 4-\gamma\right) \gamma+4\right] }$ it is a 
stable 
spiral. 
 
For the fixed point $P_{2-}$ the eigenvalues write as
\bea
&&\!\!\!\!\!\!
\mu_{1}=Z_{2-}^{*}, \nonumber \\
&&\!\!\!\!\!\!
\mu_{2,3}=\frac{1}{2}\Big\{\!-3 \left( 
1+\gamma+{{Z}_{2-}^{*}}\right)\pm \Big[9 {{\left( 
1+\gamma+{{
Z}_{2-}^{*}}\right) }^{2}}\nonumber\\
&&
\ \ \ \ \ \ \ \ \ \    \ \!
-2 {{Z}_{2-}^{*}}\left( 
12+3\gamma+4{{Z}_{2-}^{*}}\right)+\frac{9\gamma}{\alpha_{1}{{Z}_{2-}^{*}}}\Big]^{
\frac{1}{2}}\Big\},
\eea 
with 
\begin{equation}
{{Z}_{2-}^{*}}=   3\left[-\alpha_{1}- \sqrt{\alpha_{1} 
\left(\alpha_{1}-\frac{1}{6}\right)}\right]/\alpha_{1}.\nonumber
\end{equation}
Hence, this point is stable if 
 \begin{eqnarray}
&& \frac{9\gamma}{ \alpha_{1}{{Z}_{2-}^{*}}}-2{{Z}_{2-}^{*}}\left( 
12+3\gamma+4{{Z}_{2-}^{*}}\right) <0,\nonumber\\
&& -1-\gamma<{{Z}_{2-}^{*}}<0.
\end{eqnarray} 
Thus, in the physical range $0\leq\gamma<2$, 
  $P_{2-}$ 
is always 
saddle.
 
For the fixed point $P_{2+}$ the eigenvalues write as
\bea
&&
\mu_{1}=Z_{2+}^{*}, \nonumber \\
&&
\mu_{2,3}=\frac{1}{2}\Big\{\!-3 \left( 
1+\gamma+{{Z}_{2+}^{*}}\right)\pm \Big[9 {{\left( 
1+\gamma+{{
Z}_{2+}^{*}}\right) }^{2}}-\nonumber\\
&&
\ \ \ \ \ \ \ \ \ \ \ \ \ \ \! \,
2 {{Z}_{2+}^{*}}\left( 
12+3\gamma+4{{Z}_{2+}^{*}}\right)+\frac{9\gamma}{\alpha_{1}{{Z}_{2+}^{*}}}\Big]^{
\frac{1}{2}}\Big\},
\eea 
with 
\begin{equation}
{{Z}_{2+}^{*}}=3\left[-\alpha_{1}+ \sqrt{\alpha_{1} 
\left(\alpha_{1}-\frac{1}{6}\right)}\right]/\alpha_{1}.\nonumber
\end{equation}
Therefore, this solution is an attractor if it satisfies the conditions
\begin{eqnarray}
&& \frac{9\gamma}{{{Z}_{2+}^{*}}\alpha_{1}}-2{{Z}_{2+}^{*}}\left( 
12+3\gamma+4{{Z}_{2+}^{*}}\right) <0,\nonumber\\
&& -1-\gamma<{{Z}_{2+}^{*}}<0.
\end{eqnarray}  
Thus, in the physical range $0\leq\gamma<2$,
$P_{2+}$ 
is  stable for 
$\alpha_{1}>\frac{2}{3\gamma\left(4-\gamma\right)}$. 

For the fixed point $P_{3}$ the eigenvalues are given by
\bea
&&\mu_{1}=-3 \gamma,\\
&&\mu_{2,3}=-\frac{3}{2}\pm\sqrt{\frac{9}{4}-\frac{3\alpha_{2}}{2\left(\delta+\alpha_{1}
\alpha_{2}\right)}}.
\eea This critical point is an attractor if it satisfies the condition 
$\frac{\alpha_{2}}{\delta+\alpha_{1} \alpha_{2}}>0$. More specifically, it is a stable 
node for
$0<\frac{\alpha_{2}}{\delta+\alpha_{1} \alpha_{2}}<\frac{3}{2}$, while it is a stable 
spiral for 
$\frac{\alpha_{2}}{\delta+\alpha_{1} \alpha_{2}}>\frac{3}{2}$.

\section{Stability analysis of Model II: $F(T,\left(\nabla{T}\right)^2,\Box 
{T})=T+\frac{\beta_{1}\Box 
{T}}{T}+\frac{\beta_{2} \left(\Box {T}\right)^2}{T^3}+\beta_{3} e^{\frac{\sigma \Box 
{T}}{T^3}}$} 
\label{App2}

In this Appendix we investigate the stability  of Model 
II: $F(T,\left(\nabla{T}\right)^2,\Box {T})=T+\frac{\beta_{1}\Box 
{T}}{T}+\frac{\beta_{2}\left(\Box {T}\right)^2}{T^3}+\beta_{3} e^{\frac{\sigma \Box 
{T}}{T^3}}$ of subsection \ref{mod2}. The autonomous form of the system is  given in 
(\ref{Auto2}), namely
\begin{eqnarray}
&&\frac{dZ_{1}}{dN}={Z}_{1} {Z}_{2},\nonumber\\
&& \frac{dZ_{2}}{dN}={{Z}_{3}}-2 {{{Z}_{2}}^{2}},\nonumber
\end{eqnarray}
\begin{eqnarray}
\label{AutoModelII}
&& \frac{dZ_{3}}{dN}={{Z}_{4}}-3 {{Z}_{2}} {{Z}_{3}}, \nonumber\\
&&\frac{dZ_{4}}{dN}={{Z}_{5}}-4 {{Z}_{2}} {{Z}_{4}},\nonumber\\
&& \frac{d Z_{5}}{dN}={{Z}_{6}}-5 {{Z}_{2}} {{Z}_{5}},
\label{Auto2app}
\end{eqnarray}
where the function $Z_{6}(Z_{1},...,Z_{5})$ is calculated from Eqs. \eqref{SFr1} and 
\eqref{SFr2} and it reads as 
\begin{widetext}
\bea
&&
Z_{6}=
 \Bigg\{324 Z_{1}^4 \left\{\beta_{3}\sigma^2 
e^{\frac{\left[Z_{3}+Z_{2}\left(Z_{2}+3\right)\right]\sigma}
{18 Z_{1}^2}}-432 Z_{1}^6 \beta_{2}\right\}\Bigg\}^{-1}
\nonumber\\
&&
\times
\Biggl\{
\Bigg\{(6 Z_{2}^3+12 Z_{2}^2-Z_{4}+3 Z_{2} Z_{3}-3 
Z_{3})^3\sigma^4
 -18 Z_{1}^2 (Z_{4}\!-\!3 Z_{2} Z_{3}\!+\!3 Z_{3}\!-\!6 Z_{2}^3\!-\!12 Z_{2}^2) 
\Big\{3 
{{Z}_{5}}+(3\gamma- 41 {{Z}_{2}}+12){{Z}_{4}}\nonumber\\
&& \ \ \  \ \ \  \ \ \ \ \ \ \ 
-9{{Z}_{3}^{2}}
 + [ 51 {{Z}_{2}^{2}}-3 ( 
3\gamma+56){{Z}_{2}}+9\gamma+  9]
{{Z}_{3}}
+228 {{Z}_{2}^{4}}-6( 
3\gamma-61)Z_2^3-36(\gamma+1){{{Z}_{2}}^{2}}
\Big\}
\sigma^3
\nonumber\\
&& \ \ \
+324
\Big[3 Z_{1}^4 (10 Z_{2}-\gamma-2) Z_{5}
 +Z_{1}^4 (38 Z_{3}\!-\!292 Z_{2}^2
\!+\!
63\gamma Z_{2}\!+\!165 Z_{2}\!-\!18\gamma-9) Z_{4}
-3 Z_{1}^4 (47 Z_{2}-3 \gamma-
39) Z_{3}^2
\nonumber\\
&&  \ \ \ \ \ \ 
+3 Z_{1}^4 (4 Z_{2}^3\!-\!39\gamma Z_{2}^2\!-\!570 Z_{2}^2\!+\!84 \gamma 
Z_{2}\!-\!9\gamma
\!+\!
78 Z_{2}) Z_{3}
+6 Z_{1}^4 Z_{2}^2 (283 Z_{2}^3\!-\!60\gamma Z_{2}^2\!+\!383 
Z_{2}^2\!+\!18\gamma
\!-\!
99 \gamma Z_{2}-114 Z_{2})\Big]
\sigma^2
\nonumber\\
&& \ 
+17496 Z_{1}^6
\Big(4 Z_{4}-48 Z_{2} Z_{3}+
11\gamma Z_{3}+12 Z_{3}+84 Z_{2}^3 
-43 \gamma Z_{2}^2-48 Z_{2}^2+33\gamma 
Z_{2}\Big)\sigma
 +
314928 \gamma Z_{1}^8\Bigg\}  \beta_{3}\,
e^{\frac{\left[Z_{3}+Z_{2}\left(Z_{2}+3\right)\right]\sigma}{
18 Z_{1}^2}}
\nonumber\\
&&   \
-69984
\Big[6 Z_{1}^{10} (4 Z_{2}-\gamma-2) Z_{5} 
+ 2 Z_{1}^{10} (20 Z_{3}-34 Z_{2}^2+27\gamma Z_{2}+75 Z_{2}-
18\gamma-9) Z_{4}
-3 Z_{1}^{10} (12 Z_{2}-11 \gamma-42) Z_{3}^2
\nonumber\\
&& 
\ \ \ 
-
6 Z_{1}^{10} (64 Z_{2}^3\!-\!2 \gamma Z_{2}^2\!+\!102 Z_{2}^2\!-\!63\gamma Z_{2}\!-\!42 
Z_{2}\!
+\!9\gamma) Z_{3} 
+546 Z_{1}^{10} Z_{2}^5-39 (7\gamma-8) Z_{1}^{10} 
Z_{2}^4
 -
234 (\gamma+1)Z_{1}^{10} Z_{2}^3
\nonumber\\
&& 
\ \ \ 
+351\gamma Z_{1}^{10} 
Z_{2}^2\Big]
\beta_{2}
+
419904 Z_{1}^{10} \Big(2 Z_{4}\!-\!8 Z_{2} Z_{3}\!+\!6\gamma Z_{3}\!+\!6 Z_{3}\!+\!6 
Z_{2}^3
\!- \!9\gamma Z_{2}^2\!+\!18\gamma Z_{2}\Big)
\beta_{1} 
+629856 Z_{1}^{10} (2 
Z_{2}+3\gamma)\Biggr\}.
\label{ZModelIExam2}
\eea
\end{widetext}

The autonomous system \eqref{AutoModelII} 
admits six physical critical points (i.e. real and corresponding to 
 $0\leq\Omega_{DE}\leq1$), which are displayed in Table 
\ref{Table2} along with their existence conditions (without loss of generality, 
and in order to simplify the expressions, we consider $\beta_2=\frac{7 
\beta_1}{34}$). Similarly to Appendix  \ref{App1},   we perform linear 
perturbations around these critical points as $Z_{i}=Z_{i}^{*}+\delta{Z_{i}}$ and in 
this case the non-zero components of the  $5\times5$  perturbation matrix read as
{\small{
\begin{eqnarray}
&& \mathcal{M}_{11}=Z_{2},\:\:\:\:\:\:\:\:\:\:\:\:\:\:\:\:\: 
\mathcal{M}_{12}=Z_{1},\nonumber
\\
&& \mathcal{M}_{22}=-4{{Z}_{2}},\:\:\:\:\:\:\:\:\:\:\:  \mathcal{M}_{23}=1,\nonumber
\\
&& \mathcal{M}_{32}=-3{{Z}_{3}},\:\:\:\:\:\:\:\:\:\:\: \mathcal{M}_{33}=-3{{Z}_{2}}, 
\nonumber
\\
&& \mathcal{M}_{34}=1,\:\:\:\:\:\:\:\:\:\:\:\:\:\:\:\:\:\:\:\:  \mathcal{M}_{42}=-4 
{{Z}_{4}}, \nonumber
\\
&&\mathcal{M}_{44}=-4 {{Z}_{2}}, \:\:\:\:\:\:\:\:\:\:\:  \mathcal{M}_{45}=1, \nonumber
\\
&& \mathcal{M}_{51}=\frac{\partial Z_{6}}{\partial Z_{1}},\:\:\:\:\:\:\:\:\:\:\:\:\: 
\mathcal{M}_{
52}=\frac{\partial {Z}_{6}}{\partial 
{{Z}_{2}}}-5{{Z}_{5}},\nonumber
\\
&& \mathcal{M}_{53}=\frac{\partial {Z}_{6}}{\partial 
{{Z}_{3}}},\:\:\:\:\:\:\:\:\:\:\:\:\: 
\mathcal{
M}_{54}=\frac{\partial {Z}_{6}}{\partial {{Z}_{4}}},\nonumber
\\
&& \mathcal{M}_{55}=\frac{\partial {Z}_{6}}{\partial 
{{Z}_{5}}}-5 {{Z}_{2}}.
\end{eqnarray}}}
Concerning point $Q_{1}$, the corresponding eigenvalues are 
given 
by
\bea
&& \mu_{1}=-\frac{3\gamma}{2},\\
&& \mu_{2,3}=\frac{3}{4}\left(2-\gamma\right)-\sqrt{C_{+}}\pm\sqrt{C_{-}},\\
&& \mu_{4,5}=\frac{3}{4}\left(2-\gamma\right)+\sqrt{C_{+}}\pm\sqrt{C_{-}},
\eea
where we have defined the functions
\bea
&& \!\!\!\!\!\!
C_{\pm}=\frac{3\left[ 485\beta_1\pm\sqrt{7\beta_1\left( 8704- 25233\beta_1 \right) 
}\right] }{224\beta_1}.
\eea 
Although we cannot examine the sign of the real parts of the above eigenvalues 
analytically, numerically one can see that point $Q_{1}$ is always a saddle one.

In the case of fixed points $Q_{2,\pm}$ and $Q_{3\pm}$ the first eigenvalue is obviously 
$\mu_{1}=Z_{2 \pm}^{*}$. However, the other two eigenvalues are 
complicated and hence we do not give their explicit expressions here, and the 
examination of the signs of their real parts has to be done numerically. The only point 
that can be stable is $Q_{3 +}$, in the region $\beta_1>0.3$, while all the others points 
are   saddle in their respective ranges of existence. 

Finally, for the fixed point $Q_{4}$ the 
eigenvalues write as
\bea
&&\mu_{1}=-3 \gamma,\\
&& \mu_{2,3}=\frac{1}{2}\left[-3-\sqrt{D_{+}}\pm \sqrt{D_{-}}\right],\\
&& \mu_{4,5}=\frac{1}{2}\left[-3+\sqrt{D_{+}}\pm \sqrt{D_{-}}\right],
\eea where we have defined
\bea
&&\!\!\!\!\!\!\!\!\!\!\!\!\!
 D_{\pm}=\frac{-144\beta_{3}\left( 3\sigma+\beta_{1}\beta_{3}\right)  \pm 18 
E}{12{{\sigma}^{2}}
+7\beta_{1}{{\beta_{3}}^{2}}}+\frac{9}{2},\\
&&\!\!\!\!\!\!\!\!\!\!\!\!\!
E=\frac{1}{4}\left( 4 {{\sigma}^{2}}+\frac{7}{3}\beta_{1} 
{{\beta_{3}}^{2}}\right)^{1/2}
\nonumber\\
&&
\times
\Big[ 36 {{\sigma}^{2}}-576\beta_{3}\sigma+\left( 128-171\beta_{1}\right) 
{{\beta_{3}}^{2}}\Big]^{1/2}.
\eea Numerically, we find that fixed point $Q_{4}$ is always an attractor (stable node 
or stable spiral).

\end{document}